\documentclass[fleqn,usenatbib]{mnras}
\usepackage[T1]{fontenc}
\usepackage{ae,aecompl}
\usepackage{setspace}
\usepackage{natbib}
\usepackage{textcomp}
\usepackage{amsmath,amssymb,amsfonts,fullpage}
\usepackage{txfonts}
\usepackage{subfigure}
\usepackage[dvips]{graphicx}

\newcommand{\grad}{\boldsymbol{\nabla}} 
\newcommand{\di}{\partial} 

\def\Hcal{{\cal H}}

\def\Hcal{{\cal H}}  
 \def\Acal{{\cal A}}

 \title{\bf{The Contraction/Expansion History of Charon with implications for its Planetary Scale Tectonic Belt}}

\author{
Uri Malamud,$^{1}$
Hagai B. Perets,$^{1}$
Gerald Schubert$^{2}$
\\
$^{1}$Department of Physics, Technion, Israel\\
$^{2}$Department of Earth, Planetary and Space Sciences, University of California, L.A\\
}

\date{Accepted XXX. Received YYY; in original form ZZZ}

\pubyear{2016}

\begin{document}
\label{firstpage}
\pagerange{\pageref{firstpage}--\pageref{lastpage}}
\maketitle

\begin{abstract}
The New Horizons mission to the Kuiper Belt has recently revealed intriguing features on the surface of Charon, including a network of chasmata, cutting across or around a series of high topography features, conjoining to form a belt. It is proposed that this tectonic belt is a consequence of contraction/expansion episodes in the moon's evolution associated particularly with compaction, differentiation and geochemical reactions of the interior. The proposed scenario involves no need for solidification of a vast subsurface ocean and/or a warm initial state. This scenario is based on a new, detailed thermo-physical evolution model of Charon that includes multiple processes. According to the model, Charon experiences two contraction/expansion episodes in its history that may provide the proper environment for the formation of the tectonic belt. This outcome remains qualitatively the same, for several different initial conditions and parameter variations. The precise orientation of Charon's tectonic belt, and the cryovolcanic features observed south of the tectonic belt may have involved a planetary-scale impact, that occurred only after the belt had already formed.
\end{abstract}


\section{Introduction}\label{S:Intro}
\begin{figure}
\begin{center}
\includegraphics[scale=0.35]{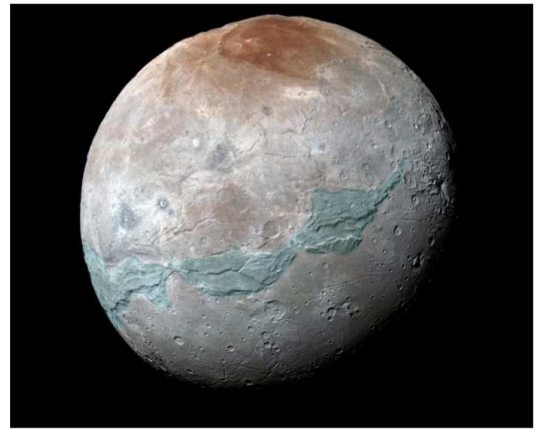}
\caption{The tectonic belt marked in light blue. Credits: NASA/JHUAPL/SwRI. Sub-spacecraft position of 25.5\textdegree N, 347.5\textdegree E and a phase angle of 38.3\textdegree. North is up.}
\label{fig:Chasmata}
\end{center}
\end{figure}

NASA's New Horizons spacecraft has recently completed a close approach to the Pluto system. The initial observations reveal Charon to have intricate geological surface features that vary in scale, shape and orientation. Among the most notable set of features, is a network of NE-SW trending fractures that cut across most of the sub-Pluto hemisphere. Perhaps the most prominent and intriguing of these, is a structure named Serenity Chasma (using informal nomenclature), a two-walled graben, several kilometers deep, and up to 60 km wide \citep{SternEtAl-2015}. In addition, the chasmata network appears to cut through or around a series of  high topography features (as seen in \cite{MooreEtAl-2016}, Figure S14), conjoining to form a belt, to which we will collectively refer to as the \textit{tectonic belt} (see the highlighted region in Figure \ref{fig:Chasmata}). This belt seems to at least partially extend to the anti-Pluto hemisphere as well \citep{BeyerEtAl-2016}, suggesting that certain compressional/extensional processes have altered the surface of Charon on a global scale during some point or points in its evolution. Serenity Chasma in particular, seems to be highly indicative of an extensional environment.   

Several alternatives that are compatible with an extensional environment can be suggested. One possible explanation is the freezing of a global sub-surface water ocean \citep{RhodenEtAl-2015}. Since the solid phase of water has a lower specific density than the liquid phase, the result would be an increase in volume. This, however, suffers from several difficulties. First, in order to account for the surface area increase that is implied by the width of Serenity Chasma (1\%, according to \cite{MooreEtAl-2016}), the sub-surface ocean must be vast (given the specific densities in Table \ref{tab:init}, a 40 km thick global ocean is required). According to our model (Section \ref{S:Results}) a vast ocean never forms because of the relatively small size of Charon and the lack of short-lived radiogenic heating. Arguably, if Charon had an initial eccentricity \citep{ChengEtAl-2014}, and if a significant amount of energy is released as a result of tidal heating during the synchronization phase of Pluto and Charon, this conclusion could change. For significant tidal heating during synchronization, Charon is required to have experienced a transient high eccentricity phase as its orbit expanded \citep{RhodenEtAl-2015}. While this is theoretically possible \citep{WardCanup-2006,ChengEtAl-2014}, the smaller size of Charon compared to Pluto makes it more likely that the tides raised by Pluto on Charon would damp Charon's eccentricity \citep{BarrCollins-2015}, leading to circularization. It is unclear if merely the initial cirularization phase could have provided the necessary tidal energy. Potentially, Charon could even have had an initial circular orbit \citep{ChengEtAl-2014}. Here, we explore a scenario that does not require tidally-driven melting of water.

Moreover, the same high eccentricity models that predict tidally-induced extensional fractures (as opposed to zero eccentricity orbital expansion with tidal bulge collapse only) require an ocean layer \citep{RhodenEtAl-2015}. However, under regular assumptions (heating by long-lived radionuclides only), the time scale of Charon's circularization (up to several Myr) is much shorter than the time required to even produce liquid water (at least 100 Myr, see Section \ref{S:Results}). Thus, in order to form tidally-induced and/or ocean freezing extensional fractures, one has to assume that a sufficient amount of heat was initially supplied in the process that led to Charon's formation. There are currently three alternative formation models for Charon: the giant impact model \citep{Canup-2005,SternEtAl-2006}, the dynamical capture model \citep{PiresDosSantosEtAl-2012} and the in-situ formation model of the Pluto-Charon binary \citep{NesvornyEtAl-2010}. None of these models can unequivocally provide the conditions necessary for the formation of an early ocean. According to \cite{Canup-2005}, the case for early melting and differentiation of Charon in the giant impact model is perhaps possible, but far weaker for Charon, compared to Pluto, given its size and the precise details concerning its formation. The analysis of \cite{PiresDosSantosEtAl-2012} does not provide any detailed information regarding the initial state of Charon after a capture. In-situ formation is the least likely model to provide this amount of heat, and therefore it remains uncertain if any of these scenarios can produce the required early ocean. 

Given these difficulties, the goal of this paper is therefore to explore an alternative model that has the potential to induce an extensional environment, without the need for a vast, pre-existing subsurface ocean, and without the necessity of a warm initial state. Such an environment may be necessary in order to explain the formation of an assemblage of global-scale intermittent chasmata that follow a similar trend (Mandjet chasma, Macross chasma and Serenity chasma, as well as possibly Argo chasma which was viewed obliquely in lower resolution images and could be a continuation of the chasmata system in the encounter hemisphere. See e.g., Figure S2 in \cite{MooreEtAl-2016}). The model also predicts compression, and we suggest that it might be related to several elevated features along the tectonic belt. The highest among these are the flanks of Serenity Chasma, and the elongated ridge extending to its right, in addition to the ridged terrain below Alice crater, which extends intermittently toward Macross chasma  (for a detailed elevation map see \cite{MooreEtAl-2016}, Figure S14). Previous papers \citep{SternEtAl-2015,MooreEtAl-2016} propose extension in order to explain the dominant chasmata, however the precise origin or cause of the elevated features along the belt and elsewhere on the surface is not discussed. Moreover, \cite{BeyerEtAl-2016} examine Serenity chasma based on lithospheric elasticity, and conclude that its observed elevated flank topography is not a flexural response. The model presented here thus provides a unique alternative that enables both extensional and compressional features in the same tectonic belt. Specifically, we predict an initial episode of compression that precedes extension, and therefore that the elevated features must pre-date the chasmata. 

The paper is arranged as follows. In Section \ref{S:Model} we outline the thermo-physical evolution model used in order to calculate the evolution of Charon. In Section \ref{S:Results} model results are presented and they are discussed in Section \ref{S:Discussion}. We conclude with a summary of our findings in Section \ref{S:Conclusions}.

\section{The Model}\label{S:Model}
The model used in this study is based on the model of \cite{MalamudPrialnik-2015}, which is an extension of earlier models by \cite{PrialnikMerk-2008} and \cite{MalamudPrialnik-2013}. This 1-dimensional code has been developed in order to study the evolving thermal and physical state of any moderate-sized icy object (large enough to be in hydrostatic equilibrium but not so large as to permit full or partial melting of the rock). It includes the following processes: (1) internal differentiation by the multiphase flow of water in a porous medium; (2) compaction by self-gravity; (3) geochemical reactions. In terms of energy sources, it considers: radiogenic heating, latent heat released/absorbed by geochemical reactions, surface insolation, gravitational energy associated with internal redistribution of mass and/or size change, and latent heat of crystallization of amorphous ice. The model treats heat transport by conduction and advection. According to the analysis by \cite{HussmanEtAl-2006}, Charon is most likely conductive. According to \cite{DeschEtAl-2009} and \cite{RubinEtAl-2014} adding parameterized convection to the calculation does not vary the results by much. Additionally they argue that the uncertainties in modelling parameterized convection, primarily in the stagnant lid regime, are at least comparable to the variations of turning off convection entirely. Thus parameterized convection has not been included in this study. 

We follow the transitions among four phases of water (amorphous ice, crystalline ice, liquid and vapour), and two phases of silicates (aqueously altered rock and non-altered rock), accounting for thermal (conductivity, heat capacity) and physical (density) changes in the solid phases as the body evolves, undergoing heating by long-lived radionuclides primarily. Internal differentiation arises from sublimation or melting of water, which triggers the multiphase flow of water inside a porous solid matrix, redistributing the internal mass. Generally, mass fluxes are proportional to the pressure gradient, or to the gradient of a function that is proportional to the pressure. When both liquid and gas phases are present, their fluxes are intricately affected by each other's presence, as described by \cite{PrialnikMerk-2008}. The mathematical modelling involves a system of coupled non-linear partial differential equations and associated initial and boundary values. It calculates mass and energy fluxes as well as the transitions among water and silicate phases. At the same time, the hydrostatic equation is solved in order to determine the density profile of the solid matrix. Thus, physical, thermal and mechanical process are coupled.

The predecessor model of \cite{MalamudPrialnik-2015} has already been used to evaluate the evolution of three moderate-sized Kuiper Belt objects (KBOs) including Charon. Nevertheless, the model presented in this study has two improvements: (1) it considers serpentine dehydration, and (2) constant solar luminosity is not assumed. The former modification is very important. The previous model only included the process of serpentinization - which leads to energy release as well as rock density decrease and absorption of water in the rock. Now the inverse process, dehydration, is also considered - leading to energy absorption as well as rock density increase and water release. The addition of dehydration considerably changes the course of Charon's evolution, not only due to the rock's physical and thermal modification, and the considerable amounts of water released, but also, since dehydration is an endothermic process, acting as a powerful energy sink that suppresses Charon's peak temperatures. The derivation of the dehydration equations is presented in \cite{MalamudPrialnik-2016}. The second modification is a small one, but it offers some improvement compared with our previous studies, as well as most evolution models, which typically consider a constant solar luminosity in order to derive the object's surface boundary condition that will define the surface temperature. In fact, the solar luminosity is not constant. It was approximately 30\% lower at the birth of the solar system and increased over time. Here we consider the change in solar luminosity as a function of time, as obtained by a 1-solar mass MESA stellar evolution \citep{PaxtonEtAl-2011}. The orbital parameters of the Pluto-Charon system on the contrary, are assumed to be constant and equal to the present day observed values, since the orbital history is harder to constrain.

\subsection{Set of Equations}\label{SS:Equations}
Considering  the transitions between four different phases of water -- amorphous ice, crystalline ice, liquid and vapour, and between two phases of silicates -- aqueously unaltered and aqueously processed, we have six different components that we denote by subscripts: $u$ - aqueously unaltered rock; $p$ - aqueously processed rock; $a$ - amorphous water ice; $c$ - crystalline water ice; $\ell$ - liquid water; $v$ - water vapour.

The independent variables are: the volume $V$; temperature $T$; densities $\rho_a$, $\rho_w=\rho_c+\rho_{\ell}$, $\rho_v$ and $\rho_d=\rho_u+\rho_p$ (the total density $\rho=\sum \rho_x$), as well as the mass fluxes $J_v$ (water vapour) and $J_\ell$ (liquid water), as functions of 1-dimensional space and time $t$.

We consider a spherically symmetric body and therefore the volume enclosed by a spherical surface of radius $r$ ($0\ \le r\ \le R$), denoted by $V$ ($0\ \le V\ \le 4\pi R^3/3$) is chosen as an independent space variable. 
Thus mass and energy fluxes will be replaced by energy or mass crossing a spherical surface per unit time, and 
$$\grad \ \Longrightarrow \frac{\di}{\di V}$$ 
The coefficients of thermal conductivity and permeability (equations \ref{eq1}, \ref{eq5} and \ref{eq6}, see below) will have to be multiplied by a factor $(4\pi r^2)^2 = (6\sqrt{\pi}V)^{4/3}$. The set of equations to be solved is:

\begin{align}
\label{eq1}
\frac{\di(\rho U)}{\di t} + \frac{\di}{\di V}\left(-K\frac{\di T}{\di V}\right)+\frac{\di(U_vJ_v + U_\ell J_\ell)}{\di V}+q_\ell\Hcal_\ell-S=0 \\
\label{eq2}
\frac{\di \rho_v}{\di t} + \frac{\di(J_v)}{\di V}  = q_v \\
\label{eq3}
\frac{\di \rho_w}{\di t} + \frac{\di J_\ell}{\di V}  = \lambda(T)\rho_a - q_v + \frac{2A_w}{A_u} \left(R_D \rho_p - R_S \rho_u\right) \\
\label{eq4}
\frac{\di \rho_a}{\di t} = -\lambda(T)\rho_a  \\
\label{eq5}
J_v = - \phi_v \frac{\di\left(P_v / \sqrt{T}\right)}{\di V}  \\
\label{eq6}
J_\ell = -\phi_\ell \left(\frac{\di(P_\ell)}{\di V}+\rho_\ell g\right) \\
\label{eq7}
Gm\rho = -4\pi (3/4\pi)^{4/3}V^{4/3}\frac{\di P}{\di V}
\end{align}

Equations (\ref{eq2}-\ref{eq4}) are the mass conservation equations, where $\lambda(T)$ is the rate of crystallization of amorphous ice, $R_S$ \citep{MalamudPrialnik-2013} is the serpentinization rate, $R_D$ \citep{MalamudPrialnik-2016} is the dehydration rate and $q_v$ is the rate of sublimation/evaporation or deposition/condensation, respectively. The mass fluxes are given by eqs (\ref{eq5}) and (\ref{eq6}), where $\phi_v$ and $\phi_\ell$ are the permeability coefficients, $P_v$ and $P_\ell$ are the vapour and liquid pressures, and $g$ is the gravitational acceleration (for the detailed multiphase flow permeability and pressure equations and their dependence on the other model parameters please refer to \cite{PrialnikMerk-2008}, equations 13-17,21-24). 

In the energy conservation equation (\ref{eq1}), $U$ denotes energy per unit mass, $\Hcal_{\ell}$ is the latent heat of fusion (melting), $q_{\ell}$ is the rate of melting/freezing, and $K$ is the effective thermal conductivity, while $(U_vJ_v + U_\ell J_\ell)$ accounts for the heat transferred by advection. The sum of all energy sources $S$ includes the energy supplied by crystallization of amorphous ice, the energy lost by sublimation, and all the other possible internal heat sources, such as radiogenic heating, tidal heating, change in gravitational potential energy and heat released or absorbed by geochemical reactions. The last equation, eq. (\ref{eq7}), imposes hydrostatic equilibrium. $m$ denotes the mass, and $G$ is the gravitational constant. The solution of the hydrostatic equation that yields the density (and hence porosity) profile, requires an equation of state (EOS), where the pressure $P$ is a function of the local density $\rho$, temperature $T$ and mass fraction of the rock. Here we use the EOS developed by \cite{MalamudPrialnik-2015} (see their equations 1-7), using identical parameters. This EOS accounts for the compaction of a porous rock/ice mixture, and is based on the best available empirical studies of ice and rock compaction, and on comparisons with rock porosities in Earth analog and Solar System silicates.

All the other variables are easily derived from the independent variables and the volume distribution. The boundary conditions adopted here are straightforward: vanishing fluxes at the centre and vanishing pressures at the surface. The surface liquid flux equals the rate of ice sublimation from the surface, whereas the surface vapour flux is the vapour that leaves the body coming from within the porous interior (outgassing). The surface heat flux is given by the balance between solar irradiation (albedo dependent), thermal emission and heat absorbed in surface sublimation of ice. We note that for an eccentric orbit, distance variations on the scale of the orbital period, would render the changes in the surface boundary condition extremely fast, and therefore the calculation extremely slow, as time steps are dynamically adjusted. For relatively small eccentricities one may circumvent this problem by considering an effective circular semi-major axis, producing an equivalent average insolation \citep{WilliamsPollard-2013}. This technique is used here for the Pluto/Charon system (the effective correction is very small, despite a notable non-circular heliocentric orbit).

\subsection{Numerical Scheme}
The model uses an adaptive-grid technique, specifically tailored for objects that change in mass or volume. Since the body is allowed to grow or shrink (as a result of various internal processes), a moving, time dependent boundary condition is implemented. The numerical solution is obtained by replacing the non-linear partial differential equations with a fully implicit difference scheme and solving a two-point boundary value problem by relaxation in an iterative process. Time steps are adjusted dynamically according to the number of iterations. In the numerical computations, where the equations are discretized in space, and the spatial grid is adapted to the varying configuration of the model, we consider a different, dimensionless space variable $x$, defined over a finite range $[c,s]$, where $c$ and $s$ are the system's boundaries (center and surface). In this case $V(x)$ must be supplied or obtained as a monotonically increasing solution of an equation that must be supplied. We normally use a very simple way of determining the spatial grid, where $V(x)$ is given by a geometric series, which keeps either the surface or the central part finely zoned, depending on the series common ratio q. Although the range of $x$ is fixed, the total volume may change with time:

\begin{equation}
V(x,t)=V_s(t)\left(1-q^{x-c}\right)/\left(1-q^{s-c}\right)
\end{equation} 

Since temporal derivatives are taken at constant $V$, whereas $V=V(x,t)$, the following transformation is implemented in the difference scheme: 

\begin{equation}
\left(\frac{\di}{\di t}\right)_V = \left(\frac{\di}{\di t}\right)_x - \left(\frac{\di V}{\di t}\right)_x \ . 
\left(\frac{\di}{\di V}\right)_t
\end{equation}

\section{Results of Evolutionary Calculation}\label{S:Results}
\subsection{Model configuration and parameters}\label{SS:Configuration}
We begin the evolution with a fully accreted object whose initial structure is homogeneous, with a well-mixed composition of rock and ice. This initial structure is the most basic to assume, since it requires no a-priori assumption of any specific formation mechanism which may lead to warm or otherwise heterogeneous accretion. By the same argument, we start with amorphous ice. Most of the important initial and physical parameters used in our model are listed in Table \ref{tab:init}. We assume an initial anhydrous rock mass fraction of 77\% corresponding to the newest mass measurement \citep{BrozovicEtAl-2015}, which yields a bulk density of 1.7 g cm$^{-3}$ \citep{SternEtAl-2015}, following a 4.6 Gyr evolution. Note that for completeness we also considered an initial rock mass fraction of 75\% based on a previous mass estimate \citep{BuieEtAl-2006}. For both selections, however, the range of the rock/ice mass ratio is compatible with a previous study \citep{MalamudPrialnik-2015}, and the resulting internal structures and evolutionary paths are qualitatively indistinguishable (see Section \ref{SS:Sensitivity}). The rock contains the long-lived radionuclides $^{235}U$, $^{40}K$, $^{238}U$ and $^{232}Th$, with initial abundances typical of meteorites. $X_0$ is the initial (anhydrous rock) mass fraction. Short-lived radionuclides can be neglected, since the accretion time of Charon is expected to be much longer than in the inner solar system \citep{KenyonBromley-2012}, on the order of tens of Myr.

\begin{table*}
\caption{Initial and physical parameters}
\centering
\smallskip
\begin{minipage}{14.2cm}
\begin{tabular}{|l|l|l|}
\hline
{\bf Parameter}                      & {\bf Symbol} & {\bf Value} \\ \hline
Initial uniform temperature          & $T_0$ & 70 K \\
Nominal $^{235}$U abundance          & $X_0$($^{235}$U) & $6.16\cdot 10^{-9}$ \\
Nominal $^{40}$K abundance           & $X_0$($^{40}$K) & $1.13\cdot 10^{-6}$ \\
Nominal $^{238}$U abundance          & $X_0$($^{238}$U) & $2.18\cdot 10^{-8}$ \\
Nominal $^{232}$Th abundance         & $X_0$($^{232}$Th) & $5.52\cdot 10^{-8}$\\
\hline
Albedo                               & $\Acal$ & 0.38 \\
Pluto-Charon semi-major axis         & $a$     & 39.264 AU\\
Pluto-Charon eccentricity            & $e$     & 0.24897 \\
Ice specific density                 & $\varrho_{a,c}$ & 0.917 g cm$^{-3}$ \\
Water specific density               & $\varrho_\ell$  & 0.997 g cm$^{-3}$ \\
Rock specific density (u)            & $\varrho_u$ &  $3.5+2.15\cdot 10^{-12}P$ g cm$^{-3}$ \\
Rock specific density (p)            & $\varrho_p$ &  $2.9+3.41\cdot 10^{-12}P$ g cm$^{-3}$ \\
Water thermal conductivity           & $K_\ell$ & $5.5\cdot 10^4$ erg cm$^{-1}$ s$^{-1}$ K$^{-1}$\\
Ice thermal conductivity (c)         & $K_c$ & $5.67\cdot 10^7/T$ erg cm$^{-1}$ s$^{-1}$ K$^{-1}$\\
Ice thermal conductivity (a)         & $K_a$ & $2.348\cdot 10^2T+2.82\cdot 10^3$\\
                                     &       & erg cm$^{-1}$ s$^{-1}$ K$^{-1}$\\
Rock thermal conductivity (u)        & $K_u$& $10^5/(0.11+3.18\cdot 10^{-4}T)+$\\
                                     &       &$3.1\cdot 10^{-5}T^3$ erg cm$^{-1}$ s$^{-1}$ K$^{-1}$\\
Rock thermal conductivity (p)        & $K_p$& $10^5/(0.427+1.1\cdot 10^{-4}T)+$\\
                                     &       &$8.5\cdot 10^{-6}T^3$ erg cm$^{-1}$ s$^{-1}$ K$^{-1}$\\
\hline
\end{tabular}
\label{tab:init}
\newline Sources: $X_0$ \citep{Prialnik-2000}; $\varrho_u$ \citep{WattAhrens-1986}; $\varrho_p$ \citep{TyburczyEtAl-1991}; $K_{\ell,c,a}$ - \cite{MalamudPrialnik-2013}; $K_{u,p}$ - \cite{MalamudPrialnik-2015}.
\end{minipage}
\end{table*}

\pagebreak[4]

\subsection{The evolutionary course}\label{SS:Evolution}
\begin{figure*}
\begin{center}
\includegraphics[scale=0.7]{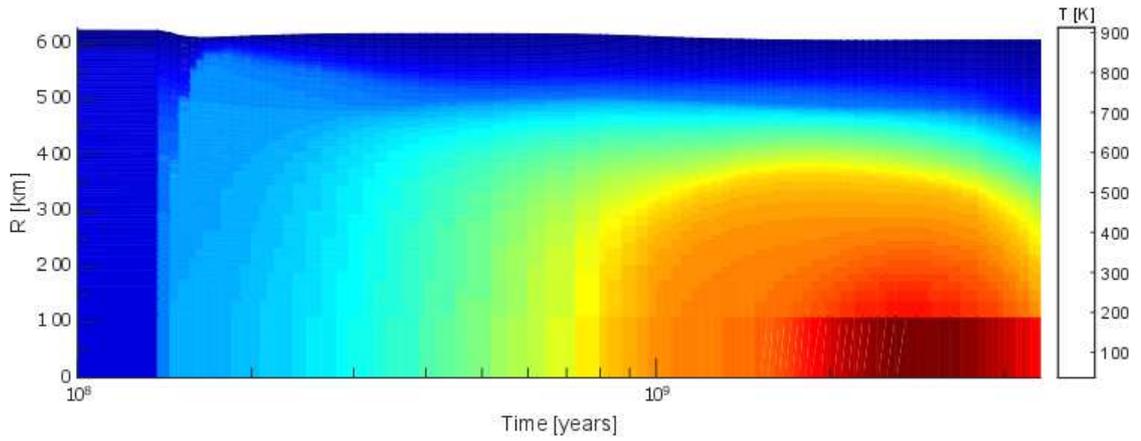}
\caption{Temperature as a function of time and radial distance.}
\label{fig:T}
\end{center}
\end{figure*}

\begin{figure*}
\begin{center}
\includegraphics[scale=0.7]{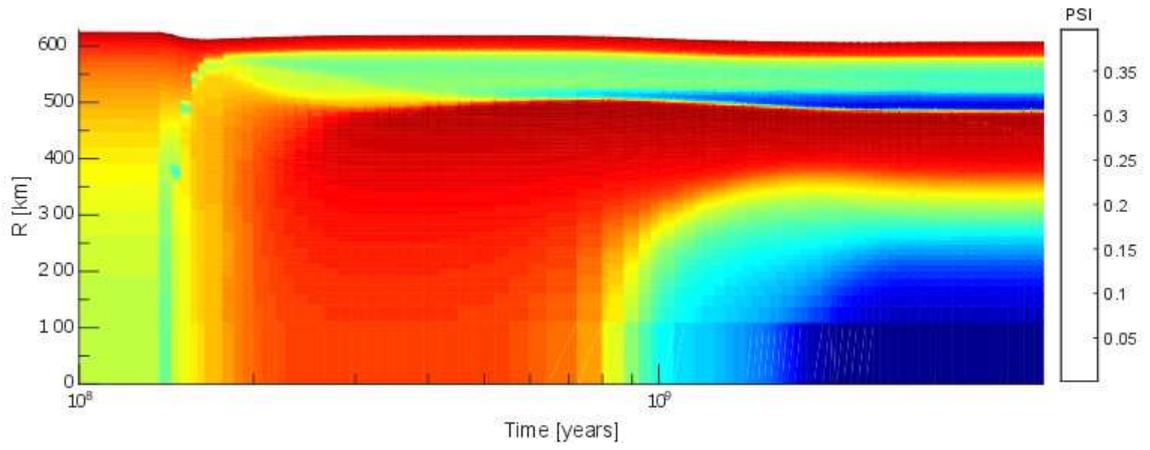}
\caption{Porosity as a function of time and radial distance.}
\label{fig:PSI}
\end{center}
\end{figure*}

The course of the evolution is illustrated by a series of figures showing properties as a function of time and radial distance from the centre of the body. Since the radius of Charon changes during evolution (initially 629 km, and eventually converging to 606 km), the upper boundary of the plots also changes with time. The evolution of Charon proceeds in several steps, corresponding to different processes in the interior:

\textbf{Step 1} (0 $< t <$ 137 Myr): In this time interval the temperatures are always below the melting point of water, as shown in Figure \ref{fig:T}, so the only process that modifies the structure (and size) of Charon is compaction of the ice in the interior. As the ice warms from an initially cold 70 K (at which point the initial bulk porosity is 28\%) it becomes more susceptible to compaction, so the radius decreases (see the porosity distribution in Figure \ref{fig:PSI} where porosity is defined as the volume fraction not occupied by solid or liquid phases). We note that the resulting radius decrease may be seen as an upper limit, since the actual specific density of ice is temperature dependent, and not constant as assumed in Table \ref{tab:init}. Instead it changes from about $\sim 0.935$ g cm$^{-3}$ at 70 K to $0.917$ g cm$^{-3}$ prior to melting. The actual radius decrease could then be reduced by up to 50-70\%. When we consider a warmer initial temperature for the ice, as in Section \ref{SS:Changes}, both compaction and ice density changes are relatively negligible.
 
\begin{figure*}
\begin{center}
\includegraphics[scale=0.7]{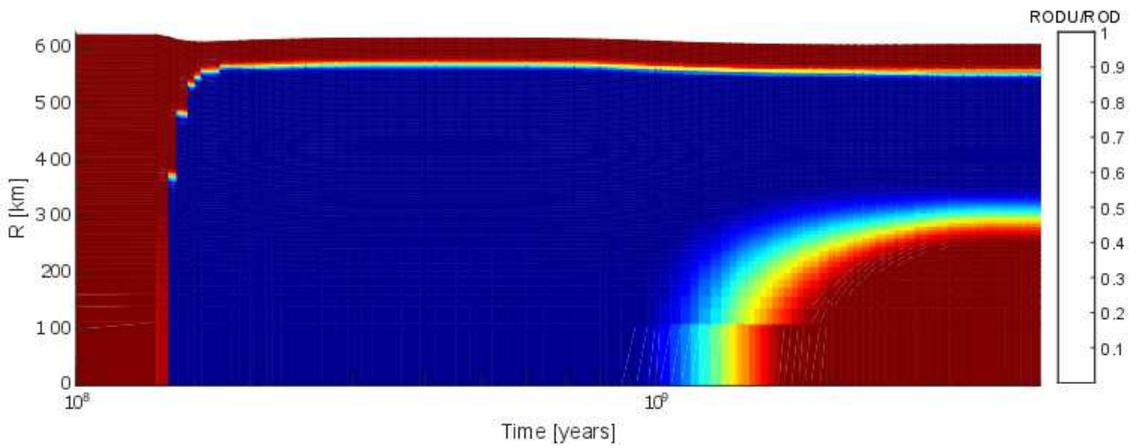}
\caption{Mass fraction of anhydrous rock (relative to total rock), as a function of time and radial distance.}
\label{fig:RODUdivROD}
\end{center}
\end{figure*}

\textbf{Step 2} (137 Myr $< t <$ 164 Myr): The melting temperature is reached. This drives rapid differentiation by liquid water flow, powered by rapid release of energy from serpentinization. In the process, the initially anhydrous rock becomes hydrated (Figure \ref{fig:RODUdivROD} shows the rock phase transitions), and a large fraction of the water in the system is absorbed in the rock. For the remainder of the paper we will refer to all water phases (including ice, liquid and vapour) not embedded onto the rock as \emph{free water}. The net effect of ice melting and incorporation in the rock is a further decrease in radius (while hydrous rock is actually more voluminous than its predecessor, the melting of ice is of greater contribution to decreasing the size). A pristine outer layer, 35 km thick, of ice/rock mixture, keeps its original anhydrous rock, since it is too cold for liquid water to reach. About two thirds of this layer (23 km) is too cold even for vapour transport, thus it retains the initial rock/ice ratio. The outermost 10 km is so cold, that it retains amorphous ice. Elsewhere the ice crystallizes. Directly beneath the pristine outer layer, there is a transition layer, from fully anhydrous rock (a relative mass fraction of 1) to fully serpentinized rock (a relative mass fraction of 0), some 25 km thick. Combined, these two layers are 60 km thick. All underlying mantle silicates are fully hydrous. Note the energetic significance of serpentinization, without which these outer layers could be much thicker (the massive amount of energy released by serpentinization, generates higher temperatures further out toward the surface).

\begin{figure*}
\begin{center}
\includegraphics[scale=0.7]{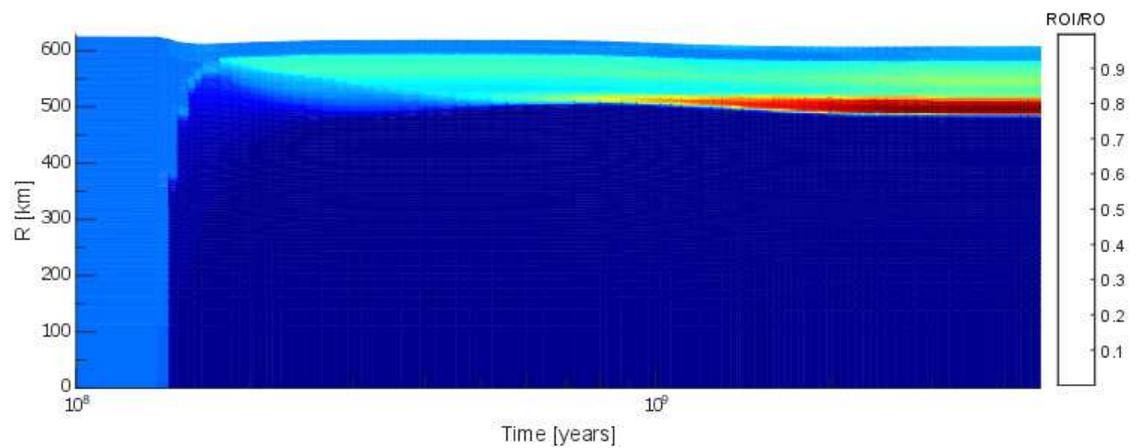}
\caption{Ice mass fraction, as a function of time and radial distance.}
\label{fig:ROIdivRO}
\end{center}
\end{figure*}

\textbf{Step 3} (164 Myr $< t <$ 450 Myr): Differentiation continues, since there is still free water (liquid) in the interior. The already hydrated rock can no longer absorb any free water, so that all remaining water is transported to the coldest regions and begins to freeze at the base of the mantle (Figure \ref{fig:ROIdivRO}), constantly thickening it, until there is essentially no longer free water in the interior by 450 Myr. At this point the ice-rich mantle is about 120 km thick, and it remains approximately that size throughout the rest of the evolution. Since the rocky core is now completely devoid of ice, its porosity has greatly increased (see the porosity distribution in Figure \ref{fig:PSI}), given Charon's pressure by self-gravity, and the (still) cold core temperatures, which do not enhance compaction in pure rock. In the absence of core compaction, the freezing of ice at the base of the mantle results in an expansion in volume, so the outcome of this stage is an increase in size. The core temperature continues to increase as a result of radiogenic heating.

\textbf{Step 4} (450 Myr $< t <$ 1 Gyr): After about 450 Myr, the temperature in the interior warms to above 450 K. This is when the rock starts to be a little more susceptible to compaction - a tendency which keeps increasing with temperature (given our compaction EOS). When core compaction is triggered, the expansion stops and the radius of Charon begins to decrease as a result of core compaction by self-gravity.

\begin{figure*}
\begin{center}
\includegraphics[scale=0.7]{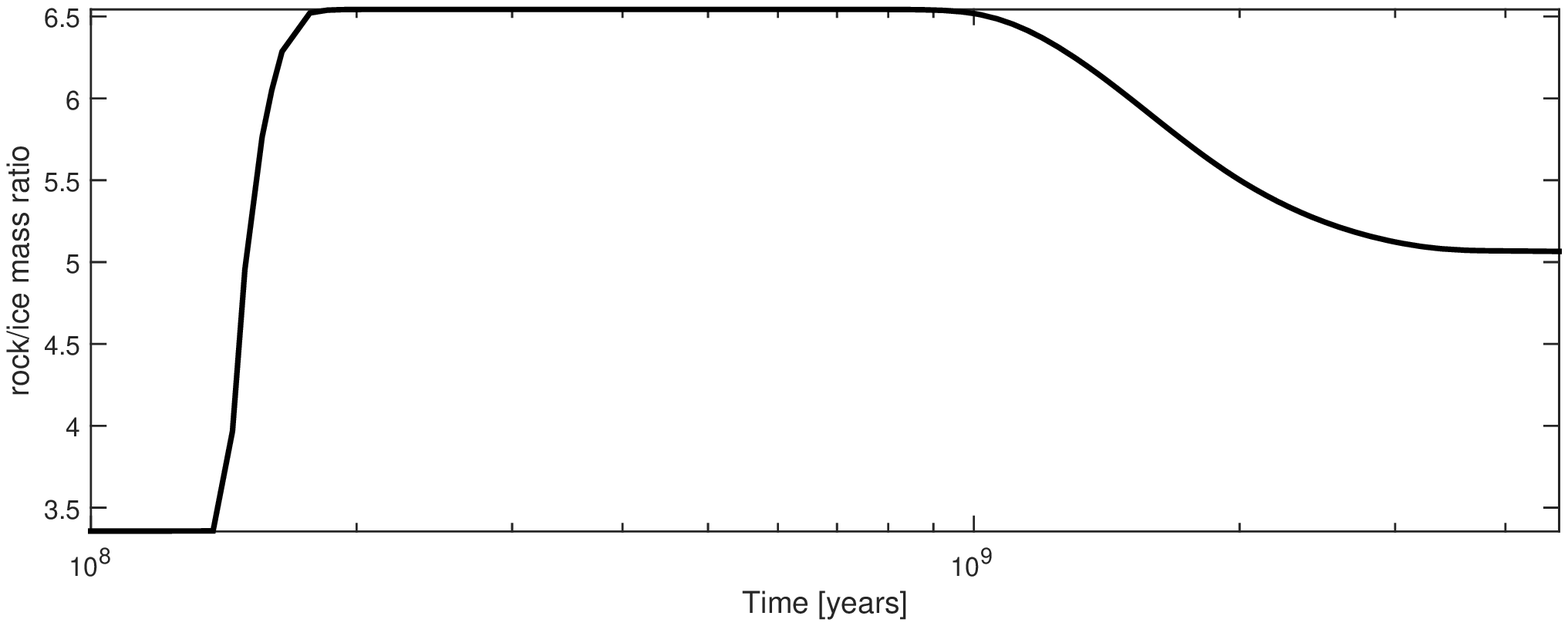}
\caption{Global rock/ice mass ratio as a function of time.}
\label{fig:RIMR}
\end{center}
\end{figure*}

\textbf{Step 5} (1 Gyr $< t <$ 4.6 Gyr): After 1 Gyr, the temperature in the core reaches approximately 675 K, at which point the reverse process to serpentinization starts increasing in rate, the rock rapidly exuding the water it had absorbed. The central part of the body becomes dehydrated, so now the stratification is more complex: an anhydrous rocky core, underlying an outer, colder, hydrous rocky layer (see Figure \ref{fig:RODUdivROD}). Dehydration also acts as a powerful internal energy sink, suppressing (but not counteracting) the increase of temperatures by radiogenic heating. As the evolution advances the inner core grows and its temperature increases. The released water migrates to the base of the mantle and freezes; so, by the end of the evolution (4.6 Gyr), the mass fraction of free water globally increases from 13.2\% (that is, after serpentinization, during which about 10\% of the water becomes embedded in the hydrated rocks) to about 16.5\%. This transition is depicted in terms of rock/ice mass ratio in Figure \ref{fig:RIMR} -- recalling that the initial rock/ice mass ratio was 3.33 (or 23 \% water mass fraction). Meanwhile, increasing core temperatures act to reduce the core porosity, hence reducing Charon's size. On the other hand, water freezing at the base of the mantle has the inverse effect. The net result of the two competing effects is that first a significant decrease in radius occurs, and only after the interior begins to cool (approximately 2.25 Gyr), when the core porosity becomes fixed at its minimum, is there a radius increase, albeit a very small one.

\begin{figure*}
\begin{center}
\includegraphics[scale=0.7]{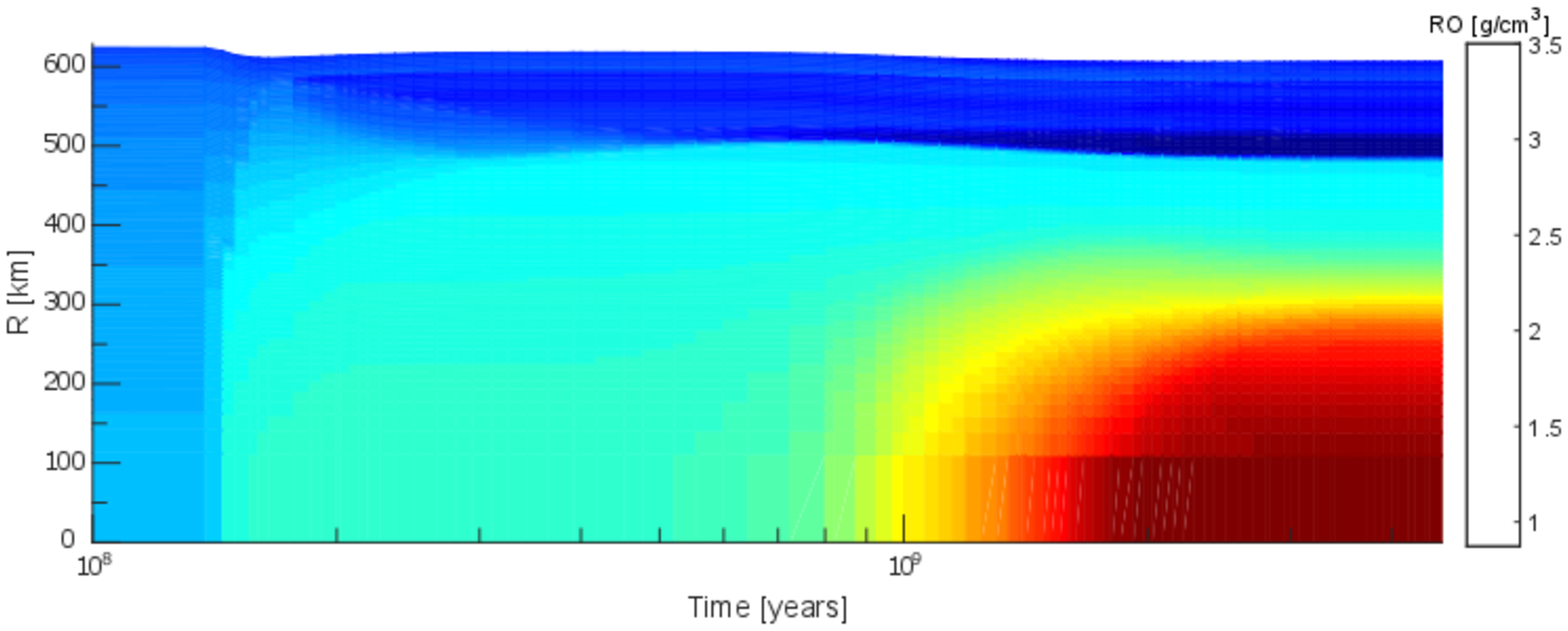}
\caption{Total density as a function of time and radial distance.}
\label{fig:RO}
\end{center}
\end{figure*}

\begin{figure}
\begin{center}
\includegraphics[scale=0.7]{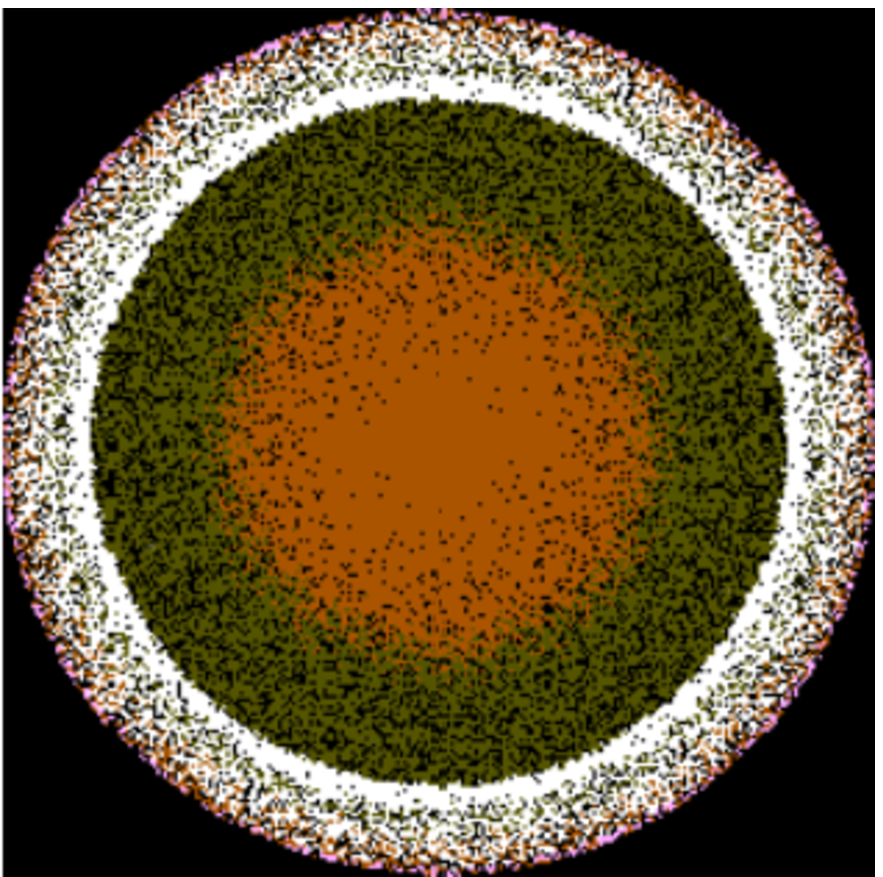}
\caption{Present day cross-section - colour interpretation: {\it black} (pores); {\it white} (crystalline ice); {\it pink} (amorphous ice); {\it brown} (anhydrous rock); and {\it olive} (hydrated rock)}.
\label{fig:CS}
\end{center}
\end{figure}

The contribution of gravitational energy by compaction/expansion or internal redistribution of mass is found to be negligible, as expected from all previous studies \citep{DeschEtAl-2009,MalamudPrialnik-2015}. At any point in time the temperature decreases monotonically from the centre of the body to the surface (Figure \ref{fig:T}). The porosity, by contrast, has two distinct peaks during most of the evolution, one near the core boundary and another in the outermost layer of the object, as shown in Figure \ref{fig:PSI}. At the centre, where the temperature and pressure are highest, the porosity is almost vanishing, increasing toward the core boundary. The same trend is exhibited in the icy mantle, which is very compact at the bottom, where the temperature and pressure are highest, and the rock fraction is lowest, in contrast with the overlying surface layers. The total density (see Figure \ref{fig:RO}) decreases monotonically from the centre of the body to the core-mantle boundary, at which point its profile becomes more complicated, with a local minimum at the base of the mantle, corresponding to the highest degree of ice enrichment (the density converges to that of pure, non-porous ice). This configuration is gravitationally unstable; however, for an overturn by Rayleigh-Taylor instability to occur, the viscosity has to be sufficiently low, and thus the temperature sufficiently high, for a suitable amount of time \citep{RubinEtAl-2014}. We find that for the bulk of the evolution presented above, mantle conditions do not permit overturn by Rayleigh-Taylor instability. Finally, a schematic depiction of the present day structure and composition of Charon, according to the model, is shown in Figure \ref{fig:CS}.

\section{Discussion}\label{S:Discussion}
\subsection{Present Day Structure Compared with Previous Models} \label{SS:Comparison}
We discuss and compare several models of Charon, focusing on its present day structure. The model presented above shows only one potential outcome of the long-term evolution of Charon, although the initial conditions and model parameters are judiciously chosen. It is important to point out, that given some changes in the parameter space (e.g., different grain specific densities, different thermal conductivities, inclusion of antifreeze to lower the water melting point, different initial assumptions like the rock/ice mass ratio or the formation temperature, etc.), within reasonable constraints, the results still remain qualitatively the same. The end structure obtained for Charon is always differentiated into a rocky core (in the above realization, it is approximately 485 km in radius) and an ice-enriched mantle (approximately 120 km thick). The rocky core is further stratified into an inner dehydrated part (250 km thick) and outer hydrated part (135 km thick), separated by a transition layer (100 km thick). The ice-enriched mantle is more complex. It is composed of pure ice at its base, the rock fraction variably increasing toward the surface. The rock undergoes a transition from fully hydrated to fully dehydrated. The outermost coldest layer preserves the original anhydrous rock, as well as the original rock/ice ratio (and it contains yet a thinner surface layer that may preserve its amorphous ice). While the sizes given above may change for different choices of parameter space, the layout remains the same, and the evolution proceeds along the same lines suggested in Section \ref{SS:Evolution}.

Other thermal evolution models have been used in order to study Charon's internal structure, reaching similar results in some instances, albeit using very different model assumptions. The model of \cite{HussmanEtAl-2006} considers the heat transport inside a zero-porosity object whose size is fixed and equals the present day observed size. Charon is assumed (not calculated) to have a 2-layered structure, an icy shell overlying a rocky core, each layer being homogeneous. It is shown to be conductive only. The core radius is 405 km, smaller than what we obtain when porosity is accounted for, although a direct comparison is inexact since their assumed size and bulk density are outdated.

The model of \cite{DeschEtAl-2009} is more sophisticated, since it also calculates the differentiation of an initially well-mixed icy object, albeit in a simpler way, assuming that rock and water separate instantaneously upon passing a certain temperature threshold. Their evolution of Charon results in a differentiated structure of a rocky core (around 420 km in radius), underlying a layer of ice, underlying an outer layer (approximately 70 km thick) that preserves the original well-mixed rock/ice material (these values vary considerably for different parameter choices). The layers are homogeneous. In a later study, \cite{RubinEtAl-2014} lower the temperature threshold, considering the effect of crustal overturn by Rayleigh-Taylor instability, to obtain only a $\sim$60 km thick outer layer, and a similarly larger rocky core (all other model parameters being equal). This configuration is somewhat closer to the final configuration obtained by our model. However, their model assumes zero porosity and layers of homogeneous density. It also does not consider geochemical reactions, thereby disregarding the important physical and energetic contributions of these processes.

To our knowledge, the most detailed thermo-physical evolution model of Charon prior to this study, is that of \cite{MalamudPrialnik-2015}. Their model explicitly considers the multiphase flow and migration of water inside a porous medium, including compaction by self-gravity, and thus follows the differentiation of Charon in greater detail. It further considers serpentinization, the transition from initially anhydrous rock to hydrous rock, accounting for the water embedded in the rock and the consequent energy release. This energy elevates the near surface temperatures, so that liquid water can reach considerably closer to the surface, altering the structure of the mantle. Their model does not however consider serpentine dehydration, which changes the physical and thermal properties of the inner rocky core, releases water back to the system, and lowers the peak core temperatures. These, as well as additional smaller modifications have been added in this study. The result is a more detailed and more complete model. Although a direct comparison to previous studies is difficult, given variations in the parameter space, we summarize in Table \ref{tab:summ} some characteristic results from all 4 studies discussed above, accentuating the qualitative differences in the present day structure.

\begin{table*}
\caption{Structure of present day Charon model via different evolution models}
\smallskip
\begin{minipage}{13.5cm}
\centering
\begin{tabular}{|c|c|c|c|c|c|c|}
\hline
Model & H06\footnote{\cite{HussmanEtAl-2006}} & D09\footnote{\cite{DeschEtAl-2009}} & R14\footnote{\cite{RubinEtAl-2014}} & MP15\footnote{\cite{MalamudPrialnik-2015}} & Section \ref{S:Results} (this study) & Units  \\ \hline
			  &     &     & 	  & 					&Anhydrous  250& \\ \cline{6-6}
Core radius& 405 & 420 & 430 & 490 			   &Transition 100& \\ \cline{6-6}
			  &     &     & 	  &   				&Hydrous	 135  & \\ \cline{1-6}
			  &     & Ice &     & Pure 10       & 20 				& \\ \cline{5-6}
Mantle	  & 200 & 110 & 110 & Enriched 35	& 40				& km\\ \cline{3-6}
thickness  &	  & Mix & 	  & Transition 40 & 25 				& \\ \cline{5-6}
			  &	  & 70  & 60  & Anhydrous  25 & 35 				& \\ \hline
Max Temp.  & --- & \multicolumn{2}{|c|}{1210-1563} & 1171   & 912 & K\\ \hline
Density	  &1.757& 1.65& 1.65& 1.63	         & 1.7          & g cm$^{-3}$\\ \hline
\end{tabular}
\label{tab:summ}
\end{minipage}
\end{table*}

The most noticeable trends, apart from increasing structural complexity with each evolution model, is (1) the larger core size, and (2) lower peak temperatures. The increase in core size is a natural outcome when accounting for non-negligible porosity and geochemical reactions, although \cite{DeschEtAl-2009} and \cite{RubinEtAl-2014} have in some cases lowered the homogeneous core density to only 2.35 g cm$^{-3}$, which may be seen as equivalent to adding constant porosity. Different peak temperatures are probably a result of different choices for the thermal conductivity coefficients, as well as the contribution of porosity, which controls (lowers) the effective thermal conductivity. Most noticeably, the low peak temperature in this study is a direct result of adding dehydration to the model, which serves as a powerful heat sink.

A common conclusion of nearly all models however, is that a near surface outer layer remains unchanged by the evolution. The main difference is in its size, which also depends on how it is defined. If it is defined as the layer overlying fully serpentinized rock, its size in this study is similar to that found by \cite{RubinEtAl-2014}, about 60 km, but if it is defined as containing only fully anhydrous rock, it is approximately half that size, 35 km. If it is defined as retaining precisely the original rock/ice ratio, it is even thinner, approximately 23 km.

Since the submission of our paper, a new paper by \cite{DeschNeveu-2017} has been published, which includes a more sophisticated calculation. The final structure at the end of their evolution, now depends on the precise timing at which Charon formed from an assumed circumplutionian disk (assumed to have formed according to the giant impact scenario), which in turn depends on the evolution of the impactors prior to the impact. Charon’s post formation temperature profile is then calculated, and its subsequent evolution includes new model assumptions and parameters. One of the new features in their code considers the notion that fine, micron-sized silicates behave differently to milimiter-sized silicate particles. The former do not separate and instead remain suspended in water liquid or ice. Different assumptions regarding the exact fraction of fines then lead to very different outcomes, which are too  numerous to be reflected in Table \ref{tab:summ}.

\subsection{Size Changes During Compaction/Expansion Episodes} \label{SS:Changes}
\begin{figure*}
\begin{center}
\includegraphics[scale=0.7]{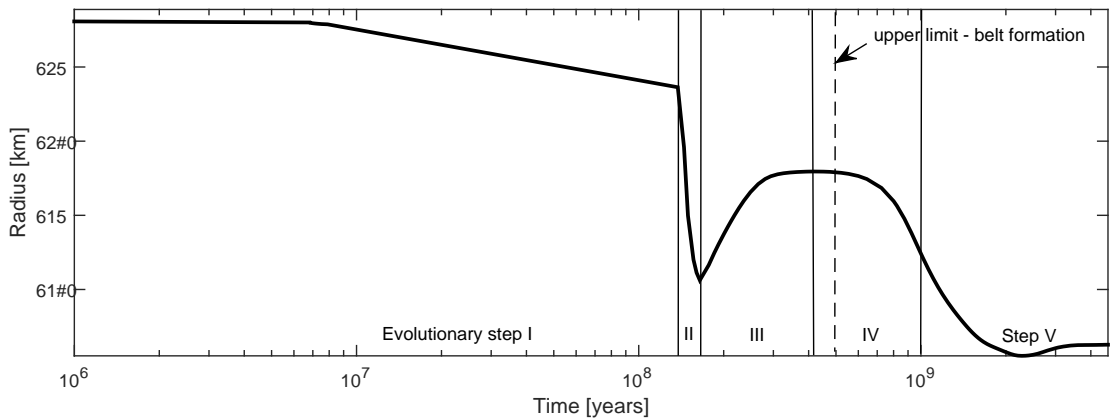}
\caption{Charon's radius as a function of time. The solid vertical lines denote the transition between various evolutionary steps: I. Ice compaction II. Melting and serpentinization III. Mantle formation IV. Core compaction V. Dehydration and freezing. The dashed vertical line marks the upper limit for the age of the belt.}
\label{fig:Radius}
\end{center}
\end{figure*}

In Section \ref{S:Results} we identify, in five evolutionary steps, processes that alter Charon's size. Our results suggest two contraction-expansion episodes, albeit on different time scales, and for different reasons. The change in radius as a function of time is shown in Fig. \ref{fig:Radius}.

Initially Charon's radius is much larger than in the present day (since it is assumed to form colder, more porous and with a well-mixed composition). The first contraction episode ($\Delta R=-18.4$ km) corresponds to a period of  compaction of warming ice followed by a period of widespread water melting, differentiation and serpentinization of initially pristine anhydrous rock (although hydrous rock is more voluminous, the net effect due to melting is a further decrease in size). Note that this radial decrease could be much smaller if Charon initially formed with higher ice temperatures, and, if the rock is initially hydrous instead of anhydrous (lowering the initial fraction of free water, in order to obtain the same overall water fraction). For example, using $T_0=220$ K we find $\Delta R$ to be merely 1/4 of the above value (for further discussion see Section \ref{SS:Sensitivity}). Once the rock is fully serpentinized, the remaining water migrates and freezes in the coldest outer parts, forming the mantle. This results in expansion ($\Delta R=7.4$ km). The second contraction episode ($\Delta R=-12.5$ km) corresponds to the contraction of the rocky core, as increasing temperatures make it more susceptible to compaction. Then the subsequent small expansion ($\Delta R=0.8$ km) results from water released by dehydration reactions in the rocky core, but it starts only after the peak temperatures in the interior begin to decrease. 

We postulate that prominent extensional fractures, such as the Serenity Chasma, could be compatible with the first expansion episode, which was exteremly rapid, on the time scale of Myr. We calculate that the additional surface area resulting from a radius increase of $\Delta R=7.4$ km would correspond to a 30 km wide fracture, forming a great circle around Charon. The width of Serenity Chasma is 50-60 km between adjacent walls. Nevertheless, Macross Chasma is no more than half that width, and the chasmata network is sporadically connected by much narrower fractures, so that on average (accounting for local variations), this radius increase could be compatible.

The chasmata in the tectonic belt are either flanked or stretched alongside elevated ridges. According to \cite{BeyerEtAl-2016,BeyerEtAl-2017}, the elevated flank topography is not likely to be a flexural response. We suggest alternatively that these ridges could have formed prior to the chasmata, by the initial contraction episode. The time scale of formation is extremely rapid, only a few tens of Myr. The precise localization of these supposed compressional features is discussed in Section \ref{SS:Orientation}. Following the initial compaction episode, the outer, cold shell, is only 1/4 its present day thickness, and it overlies warmer layers that are also composed of a rock/water mixture. It can be seen from the temperature profile in Figure \ref{fig:T} that during this time much of the water is either liquid or warm ice. Hence the effective viscosity of the underlying layers might be sufficiently low to enable isostatic subsidence of the overlying cold crust. This might considerably lower the initial elevation that resulted from the compaction (the exact compensation factor cannot be trivially calculated due to the gradient in the total density of the layers, as seen in Figure \ref{fig:RO}, but clearly the effective density difference is small). Subsidence might also weaken this part of the crust or facilitate fracture formation that may subsequently enable the focusing of extension to the same regions, as discussed in Section \ref{SS:Orientation}. Note that the extension phase starts only after the water begins to freeze (forming the mantle from the top down), as can be seen in Figure \ref{fig:ROIdivRO}. At sufficiently high latitudes, to the north of the belt, extensional features will thus not be expected to align with the main chasmata along the tectonic belt, which is consistent with \cite{BeyerEtAl-2017}.

In comparison, the second contraction episode involves a thicker mantle, and it deforms on a much longer time scale (several Gyr, 2 orders of magnitude slower). The surface might have time to relax to compensate for the slow compression of the interior. Alternatively, we might expect subtle, long-term adjustments to the reduction in surface area, accentuating existing prominent surface expressions. For example, \cite{Thomas-1988} analyzes the tectonic history of Rhea, a similar sized icy body which is also dominated by extension, however he proposes that the mega-ridges on its surface could be deformed portions of Rhea's surface that could have formed during its last evolutionary, compressional phase, according to \cite{EllsworthSchubert-1983}. While here we are considering a different internal evolution model, the geological interpretation could be similar. On Rhea, these megaridges are mostly arcuate and appear similar to the arcuate ridge around Mordor macula, for example. The final, second expansion episode is small, resulting in a very modest radius increase, so it cannot account for any of the major fractures observed on the surface. A major caveat associated with the model is that the second compaction episode cannot be reduced considerably, within reasonable parameter constraints used in our compaction equation of state. We thus emphasize the need for further studies in order to examine the long-term effects of slow compression.

\subsection{Model Sensitivity} \label{SS:Sensitivity}
The evolution results in Section \ref{SS:Evolution} and the discussion in Sections \ref{SS:Comparison} and \ref{SS:Changes} are related to only one model realization, as depicted in Section \ref{SS:Configuration}. As with any complex model, we have multiple model parameters that may be associated with uncertainties. Thus, various model realizations may lead to different outcomes. Alternatively, the same outcomes may sometimes be reached by completely different sets of parameters. Since our model is characterized by dozens of such parameters, it is implausible to fully explore the entire parameter space, while presenting dozens of different outcomes. Instead, in this section, we discuss two different choices for the initial conditions and two additional choices for important model parameters. We compare the results of these alternative model configurations to our baseline case from Section \ref{SS:Configuration}, and discuss the differences. Our baseline model will be termed \emph{Model 1}, whereas \emph{Models 2-5} are defined as follows (the title depicts the main change from baseline configuration):

\emph{Model 2 - Initial mass} We consider a change in Charon's mass. The mass that was assumed in this study \citep{BrozovicEtAl-2015} differs from the previous mass measurement \citep{BuieEtAl-2006} that was used in the predecessor model of \cite{MalamudPrialnik-2015}. The previous mass yielded a lower final bulk density at the end of the evolution, of only 1.63 g cm$^{-3}$ (versus 1.7 g cm$^{-3}$ in Section \ref{SS:Evolution}). In compatibility with the predecessor paper, we adopt the previous mass as the initial condition. Given this mass, the inferred initial rock mass fraction must be slightly higher, 77$\%$, in order to yield the observed radius at the end of the evolution.

\emph{Model 3 - Initial temperature and composition} We consider a scenario in which Charon formed with a much higher initial temperature (220 K versus only 70 K, but still below the water melting point) and with a composition of hydrated rock. This scenario might be plausible, for example, if Charon formed via the giant impact hypothesis, which might account for higher initial temperatures \citep{Canup-2005}. An initial composition of hydrated rocks is also plausible, given the likely size of the primordial Pluto or its impactor, which could have accreted from thermally processed and aqueously altered bodies, up to hundreds of km in size \citep{DavidssonEtAl-2016}. 

\emph{Model 4 - Ammonia} We consider the effect of 5$\%$ ammonia in mass fraction. If present in the internal water, ammonia can lower its melting temperature (depending on the concentration of ammonia as well as on the pressure of self-gravity, as given by \cite{Leliwa-KopystynskiEtAl-2002}). We note that ammonia has been detected on Charon's surface \citep{GrundyEtAl-2016}. In our baseline model we assume 0 $\%$ ammonia.

\emph{Model 5 - Rock grain density} We consider an alternative set of specific densities for hydrous and anhydrous rocks, 2.7 g cm$^{-3}$ and 3.25 g cm$^{-3}$ respectively (the specific density increase with pressure of self-gravity will be the same as in Table \ref{tab:init}). This choice is compatible with the sensitivity analysis performed previously by \cite{MalamudPrialnik-2015}. The specific densities chosen in the baseline model are higher, and listed in Table \ref{tab:init}. Given this new choice, the inferred initial rock mass fraction must also be set higher, 81$\%$, in order to yield the observed radius at the end of the evolution.

\begin{figure*}
	\begin{center}
		\includegraphics[scale=0.93]{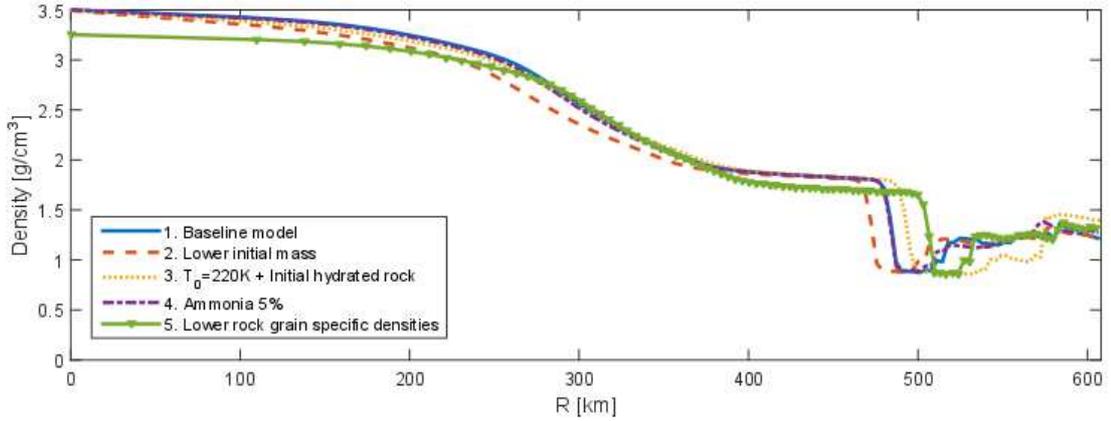}
		\caption{Charon's present-day density as a function of radial distance from the centre. The multiple plots correspond to models 1-5 with variations in initial conditions and model parameters. The steep vertical drop near a radial distance of 500 km marks the core/mantle boundary. The density at the centre equals the specific density of anhydrous rock, with zero porosity. The density at the base of the mantle is $\sim$1 g cm$^{-3}$ as in negligible-porosity water ice. }
		\label{fig:CompareRO}
	\end{center}
\end{figure*}

Since we now compare four new cases in addition to the baseline case, the level of analysis cannot be as lengthy as in Section \ref{SS:Evolution}. Instead, we set the basis for comparison after Sections \ref{SS:Comparison} and \ref{SS:Changes}. As in Section \ref{SS:Comparison}, we will first examine the final configuration obtained following the full 4.6 Gyr evolution for all five cases. Unlike in Section \ref{SS:Comparison}, here we simply plot the total density as a function of radial distance from the centre (Figure \ref{fig:CompareRO}). Although the level of detail is not as extensive as in Table \ref{tab:summ}, it provides a lot of information on both structure and composition, and in all investigated cases we obtain similar end results. The boundary between the rocky core and ice rich mantle is clearly expressed as a sharp steep drop in density. In all cases the size of the rocky core (left of the steep density decline) is approximately 500 km. The hotter inner core has become fully compressed over time and its density approaches the specific density chosen for anhydrous rock (note model 5, with a different specific density). The mantle density profile likewise has the same overall layout. In all cases the base of the mantle is the most ice-enriched and hence the least dense (despite having the lowest porosity), with the density (and rock fraction) generally increasing toward the surface. The most similar case to the baseline case is model 4. As one might expect, only the icy mantle is affected by adding ammonia, and even then the differences are small. We conclude that the end structure of Charon is more or less the same in all cases. 

\begin{figure*}
	\begin{center}
		\includegraphics[scale=0.7]{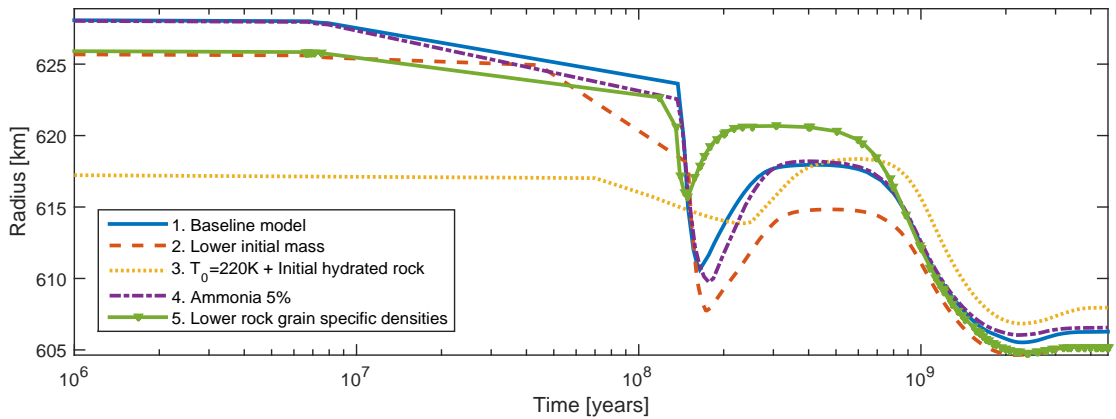}
		\caption{Charon's radius as a function of time. The multiple plots correspond to models 1-5 with variations in initial conditions and model parameters. Despite differences in radial changes, dual contraction/expansion episodes are common to all models. The models are configured such that the end radius is always similar to the present-day observed value.}
		\label{fig:CompareRadius}
	\end{center}
\end{figure*}

In terms of temporal changes, Figure \ref{fig:CompareRadius} examines the global size evolution, that is, the change in radius as a function of time. This figure is identical to Figure \ref{fig:Radius}, albeit with multiple cases plotted on the same graph. Here as well, all of our investigated cases display the same behaviour qualitatively. The overall pattern of two contraction/expansion episodes is common to all cases, and the differences are expressed in both the timing and the magnitude of each radial change. As in Figure \ref{fig:CompareRO}, model 4 displays nearly identical characteristics to the baseline model. The lower eutectic melting point as a result of adding ammonia to the model slightly enhances compaction and prolongs serpentinization, although the differences are negligible. Model 3 stands out as the most dissimilar compared to the baseline model. Here the initial temperature is already high, ice compaction is not as important, and melting is triggered earlier. Since the rock is initially hydrated, there is no energetic contribution from serpentinization to power up more rapid water melting and migration. The differentiation process is thus prolonged, and it is characterized by a lower fraction of liquid water on average, which is expressed in a more moderate radial decline. We conclude that the model is at least qualitatively insensitive to various choices of initial conditions and parameters. The highest degree of sensitivity is associated with the change in initial temperature and composition (Model 3), which can reduce the initial contraction, thus highlighting the importance of understanding Charon's formation.

\subsection{Speculation on the Belt's Localization and Orientation} \label{SS:Orientation}
In addition to Charon, many similar sized icy bodies in the inner solar system are inferred to have extensional surface features. These include the Uranian satellites Titania, Ariel, Oberon and Umbriel \citep{SmithEtAl-1986,Croft-1989}, as well as several Saturnian satellites like Tethys \citep{MooreEtAl-2004}, Rhea \citep{Thomas-1988} and Dione \citep{WagnerEtAl-2006}. In these satellites, the extensional features dominate, and many are geologically young. This partial similarity to Charon could indicate that some aspects of their formation are related, but not necessarily to tidal heating \citep{SpencerEtAl-2016}. One of the main differences is that in Charon the tectonic belt has many elevated portions. Another is that the belt is highly localized (chasmata and adjacent features are restricted to a relatively narrow band) and specifically oriented (all chasmata appear to follow a similar trend). There are very few other examples, only two icy satellites, Iapetus and Tethys, which have highly localized and specifically oriented global surface features that may also be associated with a radius change. To a lesser degree, Rhea and Dione also have similar surface features \citep{ByrneEtAl-2016}. In Iapetus' case, global compression has been one of several explanations suggested for its equatorial bulge \citep{SandwellSchubert-2010,Beuthe-2010}. In Tethys' case, global expansion has been one of the explanations suggested for Ithaca chasma, a wide extensional feature that extends approximately three-quarters of the surface \citep{MooreEtAl-2004}, due to freezing of an internal ocean. However the evolutions of both of these satellites are difficult to compare with Charon, since they are contingent upon very different initial, boundary, and external conditions (e.g., lower rock/ice mass ratio, more abundant short-lived radionuclides, warmer surface by insolation, a massive parent planet and a system of multiple massive satellites, etc.). Therefore, some hypothesis is required in order to account for the localization and orientation of Charon's tectonic belt region, as opposed to being randomly placed and randomly oriented, as one might otherwise expect.

The model presented in Section \ref{S:Model} is designed to calculate the thermo-physical evolution inside evolving icy objects, however, being 1-dimensional, the model is incapable of explaining localized phenomena. In order to calculate the precise stresses induced by the internal evolution, a different type of model is required, and therefore it is beyond the scope of this paper. Nevertheless, we suggest several possibilities which can be investigated in future studies, regarding the localization and orientation of the tectonic belt, assuming that it formed as a result of the compressional/extensional environment.

We note that although the global tectonic belt is not directly aligned with Charon's equator, it is close, and the trends of the different chasmata appear parallel, as if to suggest a shift in a primordial equator, of about 20\textdegree. Here we assume that the belt could have localized (as a result of the initial compression episode, which the model predicts to be the earliest radial change) around this primordial equator. According to one possibility, the focusing could have been made possible via thickness variations of the mantle, early in its evolution. A thinner equatorial shell could be produced due to localized tidal dissipation, as well as latitudinal differences in average solar insolation \citep{Beuthe-2010} (although in that case according to \cite{EarleBinzel-2015} the axial tilt would have had to be different compared to the present-day). \cite{SandwellSchubert-2010} also suggest that an initial axisymmetric body may provide the initial shape perturbation for subsequent buckling. Here the initial shape may be related to early tidal interaction (even simple bulge collapse, in the case that Charon was locked and had a circular orbital expansion). A precise analysis of what is required in order to focus the compression necessitates a different type of model, and is thus beyond the scope of this study. Here we simply assume that it is a possibility. 

The complete formation of the belt, however, includes the subsequent episode of extension that formed the chasmata. The focusing of extension in the same overall belt region might be related to its initial geological modifications. These may have weakened the lithosphere or created fractures/faults that made it more susceptible to the subsequent extension (see also Section \ref{SS:Changes}). A later shift in the rotational equator of Charon can be induced by a planetary-scale impact, if the impactor is sufficiently massive \citep{Safronov-1966}, leading to the currently observed orientation. This proposed impact may also be supported by several other features and terrains on the surface.

\cite{SternEtAl-2015} and \cite{MooreEtAl-2016} indicate a pronounced surface dichotomy observed between the southern and northern hemispheres of Charon, at least on the Pluto facing side (see the contrast below and above the tectonic belt in Figure \ref{fig:Chasmata} and in Figure S14 in \cite{MooreEtAl-2016} supplementary materials). The south-eastern plains (Vulcan Planum) appear significantly smoother than the rougher terrain observed on the north-western side. The age inferred for Vulcan Planum is approximately 4 Gyr, younger than the tectonic belt and the rest of the surface to its north \citep{MooreEtAl-2016,BeyerEtAl-2017}. We point out that the pronounced surface dichotomy on Charon bears resemblance to the observed surface dichotomy on Mars \citep{WilhelmsSquyers-1984}. This might also suggest a similar impact origin as used to explain Mars-dichotomy \citep{WilhelmsSquyers-1984,MarinovaEtAl-2011}. The boundary between the rough northern region and the smooth southern regions is complex, and appears to co-align with Charon's tectonic belt. Such an impact would remove a considerable volume of the outer material, redepositing it on the impact basin periphery. This loss of mass could then be fully compensated by isostatic uplift of unexcavated sub-basin material if enough energy is provided by the impact. If this material is denser than the material that was removed, a large depression could remain, like in Mars. Figure \ref{fig:RO} however shows that this must not necessarily be the case here, since the density gradient of the crust depends on the precise timing of the impact. If the impact occurred after the differentiation and formation of the tectonic belt, then the mantle density is approximately constant, and we would expect to see two distinct terrains, although the average elevation would be similar.

We emphasize that our speculations are based on high-resolution imagery that was obtained only on the sub-Pluto hemisphere. If indeed the anti-Pluto hemisphere features similar characteristics \citep{BeyerEtAl-2017}, the impact scenario has the advantage of explaining both Charon’s surface dichotomy as well as the orientation of its tectonic belt. And, as we discuss in the following Section, it may also consistently explain other Vulcan Planum features.

\subsection{Cryovolcanism} \label{SS:Cryovolcanism}
Vulcan Planum has been identified to display many features that suggest elevated temperatures, including cryovolcanic resurfacing, as well as peculiar peaks surrounded by moats \citep{MooreEtAl-2016}. \cite{DeschNeveu-2016} have interpreted the moated peaks as rapidly emplaced cryovolcanoes weighing down on and deforming the surrounding lithosphere. This flexural response demands a highly mobile layer very close to the surface, which in turn necessitates high near surface temperatures. However at Charon's distance from the sun, the temperature of its outermost layer remains permanently cold, as indicated by Figure \ref{fig:T}. This is a common result of all evolution models mentioned in Section \ref{SS:Comparison}. Even if we incorporate ammonia, a potential antifreeze (see Section \ref{SS:Sensitivity}), the crust remains cold and frozen for tens of km, and it is highly unclear how to enable the formation of these features without elevating the crust temperature. Here we suggest that a planetary scale impact is consistent with a considerable increase of crust temperature, followed by rapid cooling. 

\begin{figure*}
\begin{center}
\includegraphics[scale=0.65]{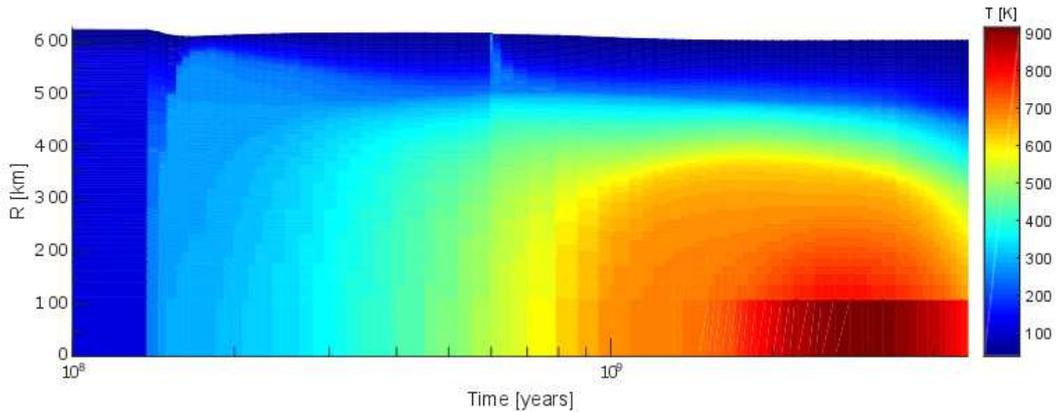}
\caption{Temperature as a function of time and radial distance, related to a massive impact.}
\label{fig:Timpact}
\end{center}
\end{figure*}

While our code is not meant to treat material excavation or uplift, to fully explore the impact scenario, we are able to explore the energetic consequence of such an impact. We show that a brief energy impulse of about $10^{26}$ J, delivered approximately 4 Gyr ago by the dichotomy-forming impact, is sufficient to bring the ice in the crust to near melting temperatures. We calculate the instantaneous release of massive amounts of energy. This is equivalent to introducing a new internal energy source, released over a short time interval (1000 s). Figure \ref{fig:Timpact} shows the resulting temperature as a function of time and radial distance. Compared with Figure \ref{fig:T}, the overall progresses of the evolution is almost identical. But during a brief period, following the impact (set arbitrarily at $\sim$600 Myr to comply with the estimation of \cite{MooreEtAl-2016}), the entire crust warms enough to support cryovolcanism. Immediately after the impact, temperatures quickly decline. The surface temperature rapidly cools on the time scale of days. Our model shows that water erupts as a result of the elevated temperatures, and Charon may lose up to 0.1\% of its mass, at a rate of $\sim 10^{15}$ g s$^{-1}$ (actual loss depends on the gas velocity and escape velocity). 4 km beneath the surface, the ice temperatures drop more slowly, merely a few degrees in approximately 50,000 yr. During this period the outgassing rate averages $\sim 10^3-10^4$ g s$^{-1}$. A further 50 Myr are required for the entire outermost 60 km to cool down to the pre-impact temperatures. 

If we decrease the amount of impact energy by one order of magnitude ($10^{25}$ J), to extend the range of possible collision energies, the resulting difference in peak surface temperature is $\Delta T\approx 40$ K, at which point the ice might be significantly colder, but still has a relatively low viscosity and high mobility, accounting for the smoothness of the surface. We also check if this energy range is compatible to first order with what is expected from the size of the impact basin. The basin size $D$ is related to the impact energy $E$ such that $D=k E^j g^u$, where the constants $k$, $j$ and $u$ are given by \cite{WilhelmsSquyers-1984} and \cite{MarinovaEtAl-2011}. The resulting energy is $\sim 5\cdot10^{25}$ J, hence this impact energy is independently consistent with our thermal interpretation, which is encouraging. Assuming a relative impact velocity of about 1 km s$^{-1}$ or slightly more \citep{DurdaStern-2000}, the mass of the impactor should be up to that of Enceladus. Assuming a relative impact velocity of up to 3 km s$^{-1}$ \citep{GreenstreetEtAl-2015}, the mass of the impactor could be an order of magnitude less than that of Enceladus. The impactor radius would thus be in the range of $\sim$100-250 km. In the current impacting environment, i.e., using estimates from today's orbital distribution and number density of Kuiper belt objects, there might be $\sim$0.1-1 such impacts over Charon's history, however the early impacting environment remains difficult to estimate \citep{GreenstreetEtAl-2015}.

\section{Conclusions}\label{S:Conclusions}
We present the main processes that have shaped Charon's long-term evolution. The evolution begins with an initially homogeneous KBO, larger than in the present, and made of a porous mixture of ice and rock. We identify two contraction-expansion episodes in Charon's history, corresponding to various processes in the interior. The present day structure we obtain is complex. The core is made of rock, with a high-density, dehydrated inner part and a more porous outer part, made of hydrated rock. The core is overlain by an ice-rich mantle, some 120 km thick. The mantle is composed of pure ice at its base, the rock fraction variably increasing toward the surface. Only the outermost coldest surface layer might retain the original composition, unaltered by the internal evolution. Compared to previous studies, we obtain a different, more stratified present-day inner structure. We also obtain a much higher rock fraction, due to the inclusion of rock hydration, as well as the contribution of porosity.

We suggest that Charon may have experienced significant changes in radius, both decrease and increase, and that these changes could be compatible with the physical appearance of its unique surface features. Our model does not require an early vast ocean, which could also potentially explain extensional surface features either by freezing and solidification of the ocean, or by high eccentricity tidally induced fractures requiring an internal liquid layer. These alternative models work only by assuming a warm initial state. In the case of this study, any formation scenario is plausible, and even the in-situ Pluto-Charon binary formation scenario, which would necessarily result in a cold initial state, could be sufficient. 

We propose that an early phase of contraction has led to the initial formation of the elevated features along the tectonic belt, followed by a subsequent phase of extension that gave rise to the chasmata. The extent of the initial radial decrease depends on Charon's initial conditions. If its temperature had been relatively warm, and its composition consisted of mainly hydrated rocks, we predict a relatively small radial decrease. If it formed cold and anhydrous, we predict a radial decrease several times larger. Even in the latter case, the high initial elevation as a result of localized contraction could be substantially reduced through isostatic subsidence, which our model predicts could have occurred before the subsequent extension.

A hypothesis is also required for the localization and specific trend of Charon's tectonic belt. Although we cannot provide a conclusive answer since our model is 1-dimensional and incapable of investigating localized phenomena, we do speculate on its origin. We hypothesize that the initial compression and extension episodes gave rise to the tectonic belt, which focused around the equatorial region. The initial focusing could have been the result of early tidal interaction with Pluto or latitudinal insolation differences. Subsequent to the formation of the tectonic belt, Charon's rotational axis must have shifted in order to account for its present day orientation. We speculate that this may be the result of a sufficiently massive impact, which could also account for Charon's strange hemispheric surface dichotomy. We show that a massive impact could further provide the energy needed in order to drive the cryovolcanism that is suggested to have occurred in the plains south of the tectonic belt. 

We conclude with a general prediction, that trans-Neptunian objects of a similar size range should exhibit significant compressional/extensional surface features. If too large (as Pluto), they might relax faster and have a warmer near surface interior. If too small, they might not even have sufficient heat for differentiation and geochemical reactions in the interior. Since variations in the short-lived radiogenic heating and the initial composition are expected to be more moderate in the Kuiper belt than in the inner solar system, so might we expect less heterogeneity in the surface features of similar sized objects in this region.

\section{Acknowledgment}\label{S:Acknowledgment}
UM and HBP acknowledge support from BSF grant number 2012384, Marie Curie FP7 career integration grant "GRAND", the Minerva center for life under extreme planetary conditions and the ISF I-CORE grant 1829/12. We would like to thank the anonymous reviewer for valuable comments and suggestions to greatly improve the quality of the paper. We would also like to thank Erez Michaeli for providing assistance in using the MESA stellar evolution code.

\newpage


\bibliographystyle{mnras} 

\begin{thebibliography}{}
	\makeatletter
	\relax
	\def\mn@urlcharsother{\let\do\@makeother \do\$\do\&\do\#\do\^\do\_\do\%\do\~}
	\def\mn@doi{\begingroup\mn@urlcharsother \@ifnextchar [ {\mn@doi@}
		{\mn@doi@[]}}
	\def\mn@doi@[#1]#2{\def\@tempa{#1}\ifx\@tempa\@empty \href
		{http://dx.doi.org/#2} {doi:#2}\else \href {http://dx.doi.org/#2} {#1}\fi
		\endgroup}
	\def\mn@eprint#1#2{\mn@eprint@#1:#2::\@nil}
	\def\mn@eprint@arXiv#1{\href {http://arxiv.org/abs/#1} {{\tt arXiv:#1}}}
	\def\mn@eprint@dblp#1{\href {http://dblp.uni-trier.de/rec/bibtex/#1.xml}
		{dblp:#1}}
	\def\mn@eprint@#1:#2:#3:#4\@nil{\def\@tempa {#1}\def\@tempb {#2}\def\@tempc
		{#3}\ifx \@tempc \@empty \let \@tempc \@tempb \let \@tempb \@tempa \fi \ifx
		\@tempb \@empty \def\@tempb {arXiv}\fi \@ifundefined
		{mn@eprint@\@tempb}{\@tempb:\@tempc}{\expandafter \expandafter \csname
			mn@eprint@\@tempb\endcsname \expandafter{\@tempc}}}
	
	\bibitem[\protect\citeauthoryear{{Barr} \& {Collins}}{{Barr} \&
		{Collins}}{2015}]{BarrCollins-2015}
	{Barr} A.~C.,  {Collins} G.~C.,  2015, \mn@doi [Icarus]
	{10.1016/j.icarus.2014.03.042}, \href
	{http://adsabs.harvard.edu/abs/2015Icar..246..146B} {246, 146}
	
	\bibitem[\protect\citeauthoryear{{Beuthe}}{{Beuthe}}{2010}]{Beuthe-2010}
	{Beuthe} M.,  2010, \mn@doi [Icarus] {10.1016/j.icarus.2010.04.019}, \href
	{http://adsabs.harvard.edu/abs/2010Icar..209..795B} {209, 795}
	
	\bibitem[\protect\citeauthoryear{{Beyer} et~al.,}{{Beyer}
		et~al.}{2016}]{BeyerEtAl-2016}
	{Beyer} R.~A.,  et~al., 2016, in Lunar and Planetary Science Conference.
	p.~2714
	
	\bibitem[\protect\citeauthoryear{{Beyer} Ross et~al.,}{{Beyer}
		et~al.}{2017}]{BeyerEtAl-2017}
	{Beyer} Ross A.,  et~al., 2017, \mn@doi [Icarus]
	{http://dx.doi.org/10.1016/j.icarus.2016.12.018}, pp~--
	
	\bibitem[\protect\citeauthoryear{{Brozovi{\'c}}, {Showalter}, {Jacobson}  \&
		{Buie}}{{Brozovi{\'c}} et~al.}{2015}]{BrozovicEtAl-2015}
	{Brozovi{\'c}} M.,  {Showalter} M.~R.,  {Jacobson} R.~A.,   {Buie} M.~W.,
	2015, \mn@doi [Icarus] {10.1016/j.icarus.2014.03.015}, \href
	{http://adsabs.harvard.edu/abs/2015Icar..246..317B} {246, 317}
	
	\bibitem[\protect\citeauthoryear{{Buie}, {Grundy}, {Young}, {Young}  \&
		{Stern}}{{Buie} et~al.}{2006}]{BuieEtAl-2006}
	{Buie} M.~W.,  {Grundy} W.~M.,  {Young} E.~F.,  {Young} L.~A.,   {Stern} S.~A.,
	2006, \mn@doi [The Astronomical Journal] {10.1086/504422}, \href
	{http://adsabs.harvard.edu/abs/2006AJ....132..290B} {132, 290}
	
	\bibitem[\protect\citeauthoryear{{Byrne}, {Schenk}, {McGovern}  \&
		{Collins}}{{Byrne} et~al.}{2016}]{ByrneEtAl-2016}
	{Byrne} P.~K.,  {Schenk} P.~M.,  {McGovern} P.~J.,   {Collins} G.~C.,  2016, in
	Enceladus and the icy moons of Saturn. Enceladus and the icy moons of Saturn.
	p.~3020
	
	\bibitem[\protect\citeauthoryear{{Canup}}{{Canup}}{2005}]{Canup-2005}
	{Canup} R.~M.,  2005, \mn@doi [Science] {10.1126/science.1106818}, \href
	{http://adsabs.harvard.edu/abs/2005Sci...307..546C} {307, 546}
	
	\bibitem[\protect\citeauthoryear{{Cheng}, {Lee}  \& {Peale}}{{Cheng}
		et~al.}{2014}]{ChengEtAl-2014}
	{Cheng} W.~H.,  {Lee} M.~H.,   {Peale} S.~J.,  2014, \mn@doi [Icarus]
	{10.1016/j.icarus.2014.01.046}, \href
	{http://adsabs.harvard.edu/abs/2014Icar..233..242C} {233, 242}
	
	\bibitem[\protect\citeauthoryear{{Croft}}{{Croft}}{1989}]{Croft-1989}
	{Croft} S.~K.,  1989, in Lunar and Planetary Science Conference.
	
	\bibitem[\protect\citeauthoryear{{Davidsson} et~al.,}{{Davidsson}
		et~al.}{2016}]{DavidssonEtAl-2016}
	{Davidsson} B.~J.~R.,  et~al., 2016, \mn@doi [Astronomy and Astrophysics]
	{10.1051/0004-6361/201526968}, \href
	{http://adsabs.harvard.edu/abs/2016A%26A...592A..63D} {592, A63}
		
		\bibitem[\protect\citeauthoryear{{Desch} \& {Neveu}}{{Desch} \&
			{Neveu}}{2016}]{DeschNeveu-2016}
		{Desch} S.~J.,  {Neveu} M.,  2016, in Lunar and Planetary Science Conference.
		p.~1647
		
		\bibitem[\protect\citeauthoryear{{Desch} \& {Neveu}}{{Desch} \&
			{Neveu}}{2017}]{DeschNeveu-2017}
		{Desch} S.,  {Neveu} M.,  2017, \mn@doi [Icarus]
		{http://dx.doi.org/10.1016/j.icarus.2016.11.037}, pp~--
		
		\bibitem[\protect\citeauthoryear{{Desch}, {Cook}, {Doggett}  \&
			{Porter}}{{Desch} et~al.}{2009}]{DeschEtAl-2009}
		{Desch} S.~J.,  {Cook} J.~C.,  {Doggett} T.~C.,   {Porter} S.~B.,  2009,
		\mn@doi [Icarus] {10.1016/j.icarus.2009.03.009}, \href
		{http://adsabs.harvard.edu/abs/2009Icar..202..694D} {202, 694}
		
		\bibitem[\protect\citeauthoryear{{Durda} \& {Stern}}{{Durda} \&
			{Stern}}{2000}]{DurdaStern-2000}
		{Durda} D.~D.,  {Stern} S.~A.,  2000, \mn@doi [Icarus]
		{10.1006/icar.1999.6333}, \href
		{http://adsabs.harvard.edu/abs/2000Icar..145..220D} {145, 220}
		
		\bibitem[\protect\citeauthoryear{{Earle} \& {Binzel}}{{Earle} \&
			{Binzel}}{2015}]{EarleBinzel-2015}
		{Earle} A.~M.,  {Binzel} R.~P.,  2015, \mn@doi [Icarus]
		{10.1016/j.icarus.2014.12.028}, \href
		{http://adsabs.harvard.edu/abs/2015Icar..250..405E} {250, 405}
		
		\bibitem[\protect\citeauthoryear{{Ellsworth} \& {Schubert}}{{Ellsworth} \&
			{Schubert}}{1983}]{EllsworthSchubert-1983}
		{Ellsworth} K.,  {Schubert} G.,  1983, \mn@doi [Icarus]
		{10.1016/0019-1035(83)90242-7}, \href
		{http://adsabs.harvard.edu/abs/1983Icar...54..490E} {54, 490}
		
		\bibitem[\protect\citeauthoryear{{Greenstreet}, {Gladman}  \&
			{McKinnon}}{{Greenstreet} et~al.}{2015}]{GreenstreetEtAl-2015}
		{Greenstreet} S.,  {Gladman} B.,   {McKinnon} W.~B.,  2015, \mn@doi [Icarus]
		{10.1016/j.icarus.2015.05.026}, \href
		{http://adsabs.harvard.edu/abs/2015Icar..258..267G} {258, 267}
		
		\bibitem[\protect\citeauthoryear{{Grundy} et~al.,}{{Grundy}
			et~al.}{2016}]{GrundyEtAl-2016}
		{Grundy} W.~M.,  et~al., 2016, \mn@doi [Science] {10.1126/science.aad9189},
		\href {http://adsabs.harvard.edu/abs/2016Sci...351.9189G} {351, aad9189}
		
		\bibitem[\protect\citeauthoryear{{Hussmann}, {Sohl}  \& {Spohn}}{{Hussmann}
			et~al.}{2006}]{HussmanEtAl-2006}
		{Hussmann} H.,  {Sohl} F.,   {Spohn} T.,  2006, \mn@doi [Icarus]
		{10.1016/j.icarus.2006.06.005}, \href
		{http://adsabs.harvard.edu/abs/2006Icar..185..258H} {185, 258}
		
		\bibitem[\protect\citeauthoryear{{Kenyon} \& {Bromley}}{{Kenyon} \&
			{Bromley}}{2012}]{KenyonBromley-2012}
		{Kenyon} S.~J.,  {Bromley} B.~C.,  2012, \mn@doi [The Astronomical Journal]
		{10.1088/0004-6256/143/3/63}, \href
		{http://adsabs.harvard.edu/abs/2012AJ....143...63K} {143, 63}
		
		\bibitem[\protect\citeauthoryear{{Leliwa-Kopysty{\'n}ski}, {Maruyama}  \&
			{Nakajima}}{{Leliwa-Kopysty{\'n}ski}
			et~al.}{2002}]{Leliwa-KopystynskiEtAl-2002}
		{Leliwa-Kopysty{\'n}ski} J.,  {Maruyama} M.,   {Nakajima} T.,  2002, \mn@doi
		[Icarus] {10.1006/icar.2002.6932}, \href
		{http://adsabs.harvard.edu/abs/2002Icar..159..518L} {159, 518}
		
		\bibitem[\protect\citeauthoryear{{Malamud} \& {Prialnik}}{{Malamud} \&
			{Prialnik}}{2013}]{MalamudPrialnik-2013}
		{Malamud} U.,  {Prialnik} D.,  2013, \mn@doi [Icarus]
		{10.1016/j.icarus.2013.04.024}, \href
		{http://adsabs.harvard.edu/abs/2013Icar..225..763M} {225, 763}
		
		\bibitem[\protect\citeauthoryear{{Malamud} \& {Prialnik}}{{Malamud} \&
			{Prialnik}}{2015}]{MalamudPrialnik-2015}
		{Malamud} U.,  {Prialnik} D.,  2015, \mn@doi [Icarus]
		{10.1016/j.icarus.2014.02.027}, \href
		{http://adsabs.harvard.edu/abs/2015Icar..246...21M} {246, 21}
		
		\bibitem[\protect\citeauthoryear{{Malamud} \& {Prialnik}}{{Malamud} \&
			{Prialnik}}{2016}]{MalamudPrialnik-2016}
		{Malamud} U.,  {Prialnik} D.,  2016, \mn@doi [Icarus]
		{10.1016/j.icarus.2015.12.046}, \href
		{http://adsabs.harvard.edu/abs/2016Icar..268....1M} {268, 1}
		
		\bibitem[\protect\citeauthoryear{{Marinova}, {Aharonson}  \&
			{Asphaug}}{{Marinova} et~al.}{2011}]{MarinovaEtAl-2011}
		{Marinova} M.~M.,  {Aharonson} O.,   {Asphaug} E.,  2011, \mn@doi [Icarus]
		{10.1016/j.icarus.2010.10.032}, \href
		{http://adsabs.harvard.edu/abs/2011Icar..211..960M} {211, 960}
		
		\bibitem[\protect\citeauthoryear{{Moore}, {Schenk}, {Bruesch}, {Asphaug}  \&
			{McKinnon}}{{Moore} et~al.}{2004}]{MooreEtAl-2004}
		{Moore} J.~M.,  {Schenk} P.~M.,  {Bruesch} L.~S.,  {Asphaug} E.,   {McKinnon}
		W.~B.,  2004, \mn@doi [Icarus] {10.1016/j.icarus.2004.05.009}, \href
		{http://adsabs.harvard.edu/abs/2004Icar..171..421M} {171, 421}
		
		\bibitem[\protect\citeauthoryear{{Moore} et~al.,}{{Moore}
			et~al.}{2016}]{MooreEtAl-2016}
		{Moore} J.~M.,  et~al., 2016, \mn@doi [Science] {10.1126/science.aad7055},
		\href {http://adsabs.harvard.edu/abs/2016Sci...351.1284M} {351, 1284}
		
		\bibitem[\protect\citeauthoryear{{Nesvorn{\'y}}, {Youdin}  \&
			{Richardson}}{{Nesvorn{\'y}} et~al.}{2010}]{NesvornyEtAl-2010}
		{Nesvorn{\'y}} D.,  {Youdin} A.~N.,   {Richardson} D.~C.,  2010, \mn@doi [The
		Astronomical Journal] {10.1088/0004-6256/140/3/785}, \href
		{http://adsabs.harvard.edu/abs/2010AJ....140..785N} {140, 785}
		
		\bibitem[\protect\citeauthoryear{{Paxton}, {Bildsten}, {Dotter}, {Herwig},
			{Lesaffre}  \& {Timmes}}{{Paxton} et~al.}{2011}]{PaxtonEtAl-2011}
		{Paxton} B.,  {Bildsten} L.,  {Dotter} A.,  {Herwig} F.,  {Lesaffre} P.,
		{Timmes} F.,  2011, \mn@doi [The Astrophysical Journal Supplement]
		{10.1088/0067-0049/192/1/3}, \href
		{http://adsabs.harvard.edu/abs/2011ApJS..192....3P} {192, 3}
		
		\bibitem[\protect\citeauthoryear{{Pires dos Santos}, {Morbidelli}  \&
			{Nesvorn{\'y}}}{{Pires dos Santos} et~al.}{2012}]{PiresDosSantosEtAl-2012}
		{Pires dos Santos} P.~M.,  {Morbidelli} A.,   {Nesvorn{\'y}} D.,  2012, \mn@doi
		[Celestial Mechanics and Dynamical Astronomy] {10.1007/s10569-012-9442-y},
		\href {http://adsabs.harvard.edu/abs/2012CeMDA.114..341P} {114, 341}
		
		\bibitem[\protect\citeauthoryear{{Prialnik}}{{Prialnik}}{2000}]{Prialnik-2000}
		{Prialnik} D.,  2000, \mn@doi [Earth Moon and Planets]
		{10.1023/A:1021577915502}, \href
		{http://adsabs.harvard.edu/abs/2000EM%26P...89...27P} {89, 27}
			
			\bibitem[\protect\citeauthoryear{{Prialnik} \& {Merk}}{{Prialnik} \&
				{Merk}}{2008}]{PrialnikMerk-2008}
			{Prialnik} D.,  {Merk} R.,  2008, \mn@doi [Icarus]
			{10.1016/j.icarus.2008.03.024}, \href
			{http://adsabs.harvard.edu/abs/2008Icar..197..211P} {197, 211}
			
			\bibitem[\protect\citeauthoryear{{Rhoden}, {Henning}, {Hurford}  \&
				{Hamilton}}{{Rhoden} et~al.}{2015}]{RhodenEtAl-2015}
			{Rhoden} A.~R.,  {Henning} W.,  {Hurford} T.~A.,   {Hamilton} D.~P.,  2015,
			\mn@doi [Icarus] {10.1016/j.icarus.2014.04.030}, \href
			{http://adsabs.harvard.edu/abs/2015Icar..246...11R} {246, 11}
			
			\bibitem[\protect\citeauthoryear{{Rubin}, {Desch}  \& {Neveu}}{{Rubin}
				et~al.}{2014}]{RubinEtAl-2014}
			{Rubin} M.~E.,  {Desch} S.~J.,   {Neveu} M.,  2014, \mn@doi [Icarus]
			{10.1016/j.icarus.2014.03.047}, \href
			{http://adsabs.harvard.edu/abs/2014Icar..236..122R} {236, 122}
			
			\bibitem[\protect\citeauthoryear{{Safronov}}{{Safronov}}{1966}]{Safronov-1966}
			{Safronov} V.~S.,  1966, Soviet Astronomy, \href
			{http://adsabs.harvard.edu/abs/1966SvA.....9..987S} {9, 987}
			
			\bibitem[\protect\citeauthoryear{{Sandwell} \& {Schubert}}{{Sandwell} \&
				{Schubert}}{2010}]{SandwellSchubert-2010}
			{Sandwell} D.,  {Schubert} G.,  2010, \mn@doi [Icarus]
			{10.1016/j.icarus.2010.06.025}, \href
			{http://adsabs.harvard.edu/abs/2010Icar..210..817S} {210, 817}
			
			\bibitem[\protect\citeauthoryear{{Smith} et~al.,}{{Smith}
				et~al.}{1986}]{SmithEtAl-1986}
			{Smith} B.~A.,  et~al., 1986, \mn@doi [Science] {10.1126/science.233.4759.43},
			\href {http://adsabs.harvard.edu/abs/1986Sci...233...43S} {233, 43}
			
			\bibitem[\protect\citeauthoryear{{Spencer} et~al.,}{{Spencer}
				et~al.}{2016}]{SpencerEtAl-2016}
			{Spencer} J.~R.,  et~al., 2016, in Lunar and Planetary Science Conference.
			p.~2440
			
			\bibitem[\protect\citeauthoryear{{Stern} et~al.,}{{Stern}
				et~al.}{2006}]{SternEtAl-2006}
			{Stern} S.~A.,  et~al., 2006, \mn@doi [Nature] {10.1038/nature04548}, \href
			{http://adsabs.harvard.edu/abs/2006Natur.439..946S} {439, 946}
			
			\bibitem[\protect\citeauthoryear{{Stern} et~al.,}{{Stern}
				et~al.}{2015}]{SternEtAl-2015}
			{Stern} S.~A.,  et~al., 2015, \mn@doi [Science] {10.1126/science.aad1815}, 350
			
			\bibitem[\protect\citeauthoryear{{Thomas}}{{Thomas}}{1988}]{Thomas-1988}
			{Thomas} P.~G.,  1988, \mn@doi [Icarus] {10.1016/0019-1035(88)90121-2}, \href
			{http://adsabs.harvard.edu/abs/1988Icar...74..554T} {74, 554}
			
			\bibitem[\protect\citeauthoryear{{Tyburczy}, {Duffy}, {Ahrens}  \&
				{Lange}}{{Tyburczy} et~al.}{1991}]{TyburczyEtAl-1991}
			{Tyburczy} J.~A.,  {Duffy} T.~S.,  {Ahrens} T.~J.,   {Lange} M.~A.,  1991,
			\mn@doi [Journal of Geophysical Research: Solid Earth] {10.1029/91JB01573},
			96, 18011
			
			\bibitem[\protect\citeauthoryear{{Wagner}, {Neukum}, {Giese}, {Roatsch},
				{Wolf}, {Denk}  \& {Cassini Iss Team}}{{Wagner}
				et~al.}{2006}]{WagnerEtAl-2006}
			{Wagner} R.,  {Neukum} G.,  {Giese} B.,  {Roatsch} T.,  {Wolf} U.,  {Denk} T.,
			{Cassini Iss Team} 2006, in {Mackwell} S.,  {Stansbery} E.,  eds,  Lunar and
			Planetary Inst.~Technical Report Vol. 37, 37th Annual Lunar and Planetary
			Science Conference.
			
			\bibitem[\protect\citeauthoryear{{Ward} \& {Canup}}{{Ward} \&
				{Canup}}{2006}]{WardCanup-2006}
			{Ward} W.~R.,  {Canup} R.~M.,  2006, \mn@doi [Science]
			{10.1126/science.1127293}, \href
			{http://adsabs.harvard.edu/abs/2006Sci...313.1107W} {313, 1107}
			
			\bibitem[\protect\citeauthoryear{{Watt} \& {Ahrens}}{{Watt} \&
				{Ahrens}}{1986}]{WattAhrens-1986}
			{Watt} J.~P.,  {Ahrens} T.~J.,  1986, \mn@doi [Journal of Geophysical Research:
			Solid Earth] {10.1029/JB091iB07p07495}, 91, 7495
			
			\bibitem[\protect\citeauthoryear{{Wilhelms} \& {Squyres}}{{Wilhelms} \&
				{Squyres}}{1984}]{WilhelmsSquyers-1984}
			{Wilhelms} D.~E.,  {Squyres} S.~W.,  1984, \mn@doi [Nature] {10.1038/309138a0},
			\href {http://adsabs.harvard.edu/abs/1984Natur.309..138W} {309, 138}
			
			\bibitem[\protect\citeauthoryear{{Williams} \& {Pollard}}{{Williams} \&
				{Pollard}}{2002}]{WilliamsPollard-2013}
			{Williams} D.~M.,  {Pollard} D.,  2002, \mn@doi [International Journal of
			Astrobiology] {10.1017/S1473550402001064}, \href
			{http://adsabs.harvard.edu/abs/2002IJAsB...1...61W} {1, 61}
			
			\makeatother
		\end{thebibliography}

\bsp	
\label{lastpage}
\end{document}